\documentclass{aa}
\usepackage{graphicx}
\usepackage{times}
\usepackage{mathptm}
\fontfamily{ptmcm}\selectfont
\begin{document}

\thesaurus{04.19.1; 13.18.1; 04.03.1}
	   
\title{The ATESP Radio Survey}

\subtitle{II. The Source Catalogue}

\author{I. Prandoni \inst{1,2}
	\and L. Gregorini \inst{3,2}
	\and P. Parma \inst{2}
	\and H.R. de Ruiter \inst{4,2}
	\and G. Vettolani \inst{2}
	\and M.H. Wieringa \inst{5}
	\and R.D. Ekers \inst{5}
	}
\offprints{I. Prandoni}
\mail{prandoni@ira.bo.cnr.it}

\institute{Dipartimento di Astronomia, Universit\`a di Bologna, via Ranzani 1,
I--40126, Bologna, Italy
\and Istituto di Radioastronomia, CNR, Via\,Gobetti 101, I--40129, 
Bologna, Italy
\and Dipartimento di Fisica, Universit\`a di Bologna, Via Irnerio 46,
I--40126, Bologna, Italy
\and Osservatorio Astronomico di Bologna, Via Ranzani 1, I--40126, 
Bologna, Italy
\and Australia Telescope National Facility, CSIRO, P.O. Box 76, Epping, 
NSW2121, Australia
}

\date{Received 08 March 2000 / Accepted 23 June 2000}

\titlerunning{The ATESP Radio Survey. II}
\authorrunning{I. Prandoni et al.}

\maketitle
\begin{abstract}
This paper is part of a series reporting the results 
of the \emph{Australia Telescope ESO Slice Project} (ATESP) radio survey 
obtained at 1400 MHz with the \emph{Australia Telescope 
Compact Array} (ATCA) over the region covered by the \emph{ESO Slice Project}
(ESP) galaxy redshift survey. 
The survey consists of 16 radio mosaics with $\sim 8\arcsec \times 14\arcsec$
resolution and uniform sensitivity ($1 \sigma$ noise 
level $\sim$79~$\mu$Jy) over the whole area of the ESP redshift survey 
($\sim 26$ sq.~degrees at $\delta \sim -40\degr$). \\ 
Here we present the catalogue derived from the ATESP survey. 
We detected $2960$ distinct radio sources down to a flux density limit of 
$\sim 0.5$ mJy ($6\sigma$), 1402 being sub-mJy sources.
We describe in detail the procedure followed for the source extraction and 
parameterization. The internal accuracy of the source parameters was tested
with Monte Carlo simulations and possible systematic effects (e.g. bandwidth 
smearing) have been quantified. \\
 

\end{abstract}

\section{Introduction}\label{sec-intr}

Recent deep radio surveys ($S << 1$ mJy) have shown that normalized radio 
source counts flatten below a few mJy. This has been interpreted as being 
due to a new population which does not show up at higher fluxes where counts 
are dominated by active galactic nuclei (AGN). 
To clarify its nature it is necessary to
get detailed information on the radio properties of normal galaxies in the 
nearby universe (z $\la$ 0.15), down to faint flux limits and to have at 
hand large samples 
of mJy and sub-mJy sources, for subsequent optical identification and 
spectroscopic work. To this end we have
surveyed a large area ($\sim $ 26 sq. degr) with the ATCA at 1.4 GHz with a 
bandwidth of $2\times128$ MHz. The 
properties of normal
nearby galaxies can be easily derived, because this area coincides
with the region in which the ESP redshift survey was conducted 
(Vettolani et al. \cite{Vettolani97}). Samples of faint galaxies over large 
areas are 
necessary to avoid bias due to local variations in their
properties. Present samples of faint radio sources are confined to small 
regions with insufficient source statistics.    \\

The present paper contains the radio source catalogue derived from the ATESP
survey. The full outline of the radio survey,
its motivation in comparison with other surveys, and the description of 
the mosaic observing 
technique which was used to obtain an optimal combination of uniform and high 
sensitivity over the whole area has been presented in Prandoni et 
al. (2000, Paper I). The source counts derived from the ATESP survey will be 
presented and discussed separately in a following paper.
The paper is organized as follows: 
in Sect.~\ref{sec-det} we describe the source detection and parameterization;
the catalogue is described in Sect.~\ref{sec-cat} and the 
accuracy of the parameters (flux densities, sizes and positions) are 
discussed in Sect.~\ref{sec-errors}. 

\section{Source Detection and Parameters}\label{sec-det}

The ATESP survey consists of 16 
radio mosaics with spatial resolution 
$\sim 8\arcsec \times 14\arcsec$. The survey was 
designed so as to provide uniform sensitivity over the whole region 
($\sim 26$ sq.~degrees) of the ESP redshift survey. 
To achieve this goal a larger
area was observed, but we have excluded from the analysis the external  
regions (where the noise is not uniform and increases radially).
In the region with uniform sensitivity the noise level
varies from 69 $\mu$Jy to 88
$\mu$Jy, depending on the radio mosaic, with an average of 79 $\mu$Jy
(see $\sigma_{\rm fit}$ values reported in Table~3 of paper I and repeated
also in Table~\ref{tab-mospar} of Appendix~\ref{sec-appb}, at the end of 
this paper). For consistency
with paper I, in the following such
sensitivity average values are denoted by the symbol $\sigma_{\rm fit}$. 
This means
that sensitivity values have been defined as the \emph{full width at 
half maximum} (FWHM) of the Gaussian that fits the pixel flux density 
distribution in each mosaic (see paper I for more details). \\
A number of source detection and parameterization algorithms are
available, which were developed for deriving catalogues of components 
from radio surveys. We decided to use the algorithm  
\emph{Image Search and Destroy} (IMSAD) available as part of the 
\emph{Multichannel Image Reconstruction, Image Analysis and Display} (MIRIAD) 
package 
(Sault \& Killeen \cite{Sault95}), as it is particularly suited to images 
obtained with the ATCA. \\
IMSAD selects all the connected regions of pixels (\emph{islands}) 
above a given flux 
density threshold. The \emph{islands} are the sources (or the source 
components) present in the image above a certain flux limit. Then IMSAD 
performs a two-dimensional
Gaussian fit of the \emph{island} flux distribution and 
provides the following parameters: position of the 
centroid (right ascension, $\alpha$, and declination, $\delta$), 
peak flux density ($S_{\rm peak}$), integrated flux density ($S_{\rm total}$), 
fitted angular size (major, $\theta_{\rm maj}$, and minor, $\theta_{\rm min}$, 
FWHM axes, not deconvolved for the beam) and position angle (P.A.). \\
IMSAD proved to have an average 
success rate of $\sim 90\%$ down to very faint flux levels (see below). 
Since IMSAD attempts to fit a single Gaussian to each \emph{island}, it 
obviously tends to fail (or to provide very poor source parameters) 
when fitting complex (\emph{i.e.} non-Gaussian) shapes. 

\subsection{Source extraction}

We used IMSAD to extract and parameterize 
all the sources and/or components in the uniform sensitivity region of 
each mosaic\footnote{To avoid interpolation and completeness problems 
the effective search 
area in each mosaic was slightly larger both in declination and in 
right ascension. The masked region in mosaic fld69to75 
has been excluded from the search (see paper I for more details).}.
As a first step, a preliminary list containing all detections
with $S_{\rm peak}\geq 4.5\sigma$ (where $\sigma$ is the average mosaic 
rms flux density) was extracted. Detection thresholds vary from 0.3 mJy
to 0.4 mJy, depending on the radio mosaic.

\subsection{Inspection}\label{sec-insp} 

We visually inspected all \emph{islands} ($\sim 5000$) detected, in order 
to check for obvious failures and/or possibly poor parameterization, that 
need further analysis.
Problematic cases were classified as follows: 
\begin{itemize}
\item 
\emph{islands} for which the automated algorithm provides
poor parameters and therefore are to be re-fitted with a single Gaussian,
constraining in a different way the initial conditions 
($\sim 5\%$ of the total). 
\item 
\emph{islands} that could be better described by two or more split 
Gaussian components ($\sim 3\%$ of the total);  
\item 
\emph{islands} which cannot be described by a single or multiple Gaussian 
fit ($\sim 1\%$ of the total); 
\item 
\emph{islands} corresponding to obviously unreal sources,
typically noise peaks and/or image artifacts in noisy regions of the images 
($\sim 2\%$ of the total); 
\end{itemize}
The goodness of Gaussian fit parameters was checked by comparing them with 
reference values, defined as follows.
Positions and peak flux densities
were compared to the corresponding values derived by a 
second-degree interpolation of the \emph{island}. Such interpolation 
usually provides very accurate positions and peak fluxes.
Gaussian integrated fluxes were compared to the ones derived directly by 
 summing pixel per pixel the flux density in the source area, 
defined as the region enclosed by the $\geq 3\sigma$ flux density contour.
Flux densities were considered good whenever the difference between the 
Gaussian and the reference value $S-S_{\rm ref}$ was $\leq 0.2 S_{\rm ref}$.
Positions were considered good whenever they did fall within the 
$0.9 S_{\rm peak}$ flux contour.

\begin{figure}[t]
\resizebox{\hsize}{!}{\includegraphics{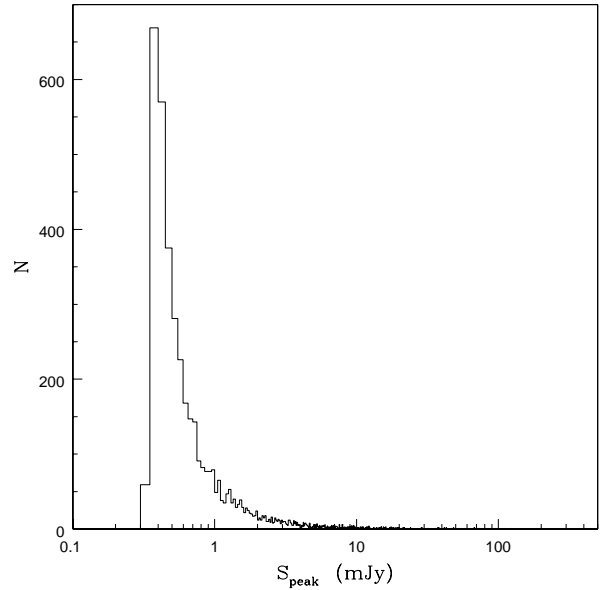}}
\caption{Peak flux density distribution for all the ATESP radio sources 
(or source components) with $S_{\rm peak} \geq 4.5\sigma$.
\label{fig-fluxhist}}
\end{figure}

\subsection{Re-fitting}\label{sec-refit}

For the first three groups listed above 
\emph{ad hoc} procedures were attempted aimed at improving the fit. \\
Single-component fits were considered satisfactory 
whenever positions and flux densities satisfy the tolerance criteria 
defined above.\\
In a few cases Gaussian fits were able to provide good values for positions 
and peak flux densities, but did fail in 
determining the integrated flux densities. This happens typically at faint 
fluxes ($<10\sigma$). 
Gaussian sources with a poor $S_{\rm total}$ value are 
flagged in the catalogue (see Sect.~\ref{sec-cformat}).    \\
The \emph{islands} successfully split in two or three components 
are 67 in total (64 with two 
components and 3 with three components).    \\
For non-Gaussian sources we adopted as parameters the reference 
positions and flux densities defined above. The source position angle 
was determined by 
the direction in which the source is most extended and the source axes 
were defined as \emph{largest angular sizes} (\emph{las}), i.e. the maximum 
distance between two opposite points belonging to the $3\sigma$ 
flux density 
contour along (major axis) and perpendicular to (minor axis) the same 
direction. 
All the non-Gaussian sources are flagged in the catalogue 
(see Sect.~\ref{sec-cformat}). 

\begin{figure}[t]
\resizebox{\hsize}{!}{\includegraphics{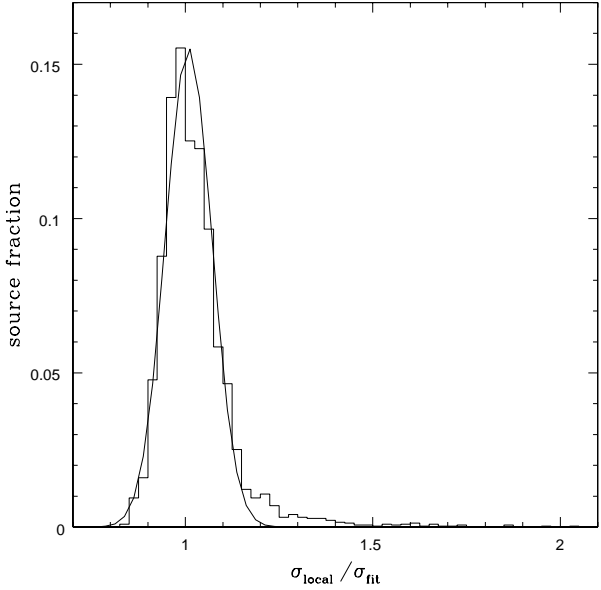}}
\caption{Local to average noise ratio distribution for all the sources with 
$S_{\rm peak} \geq 6\sigma_{\rm fit}$. Local noise is defined 
as the average noise in a $\sim 8\arcmin$ sided box around each source. 
The distribution is well fitted by a Gaussian with ${\rm FWHM} = 0.14$ and 
peak value equal to 1.01 (solid line). 
The excess at large $\sigma_{\rm local}/\sigma_{\rm fit}$ values
is due to the presence of systematic noise effects (see text).
\label{fig-lrmshist}}
\end{figure}

\begin{figure}[t]
\resizebox{\hsize}{!}{\includegraphics{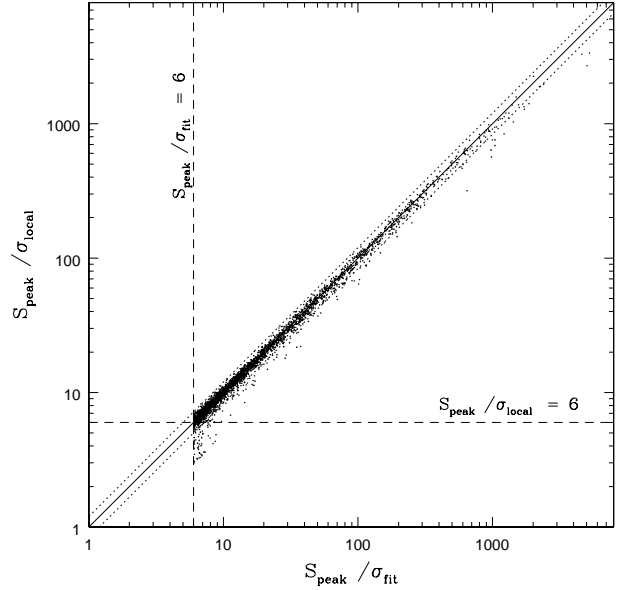}}
\caption{$S_{\rm peak}/\sigma_{\rm local}$ vs. $S_{\rm peak}/\sigma_{\rm fit}$.
As expected for random noise distributions, the two signal-to-noise ratios 
show a tight correlation. Dashed lines indicate the 
$S_{\rm peak}=6\sigma_{\rm fit}$ (vertical line) and 
$S_{\rm peak}=6\sigma_{\rm local}$ 
(horizontal line) cut-off respectively. Also shown are the lines (dotted) 
between which, under the assumption of a normal noise distribution,
one would find 99.9\% of the sources. 
\label{fig-lrms}}
\end{figure}

\section{The Source Catalogue}\label{sec-cat}

The procedures described in the previous section yielded a preliminary list
of sources (or source components) for further investigation. In order to 
minimize the incompleteness effects 
present at fluxes approaching the source extraction threshold 
(see Fig.~\ref{fig-fluxhist}) we decided to insert in the 
final catalogue only the sources  with $S_{\rm peak} \geq 6\sigma$, where 
$\sigma$ is the mosaic rms flux density. 
This threshold has been chosen after inspection of the local noise 
distribution. The local noise ($\sigma_{\rm local}$) has been defined 
as the average noise value in a box of about $8\arcmin \times 8\arcmin$ size 
around a source. 
Usually the local noise does not show significant systematic 
departures from the mosaic average rms value: the 
$\sigma_{\rm local}/ \sigma_{\rm fit}$ distribution can be 
described fairly well by a Gaussian with ${\rm FWHM} = 0.14$ and peak value 
equal to 1.01 (see Fig.~\ref{fig-lrmshist}). \\
This can be seen also in Fig.~\ref{fig-lrms}, where we show, for each source, 
the signal-to-noise 
ratio defined using both $\sigma_{\rm local}$ and $\sigma_{\rm fit}$. 
The two signal-to-noise ratios 
mostly agree with each other, although 
a number of significant departures are evident for the 
faintest sources. 
This is due to the presence of some residual areas  
where the noise is not random due to 
systematic effects (noise peaks and stripes). This is caused by the
limited dynamical range in presence of very strong sources 
(stronger than $50-100$ mJy, see also paper I).  
It is worth noting that also the systematic departures from the expected 
behavior at the brightest end of the plot ($S_{\rm peak} / \sigma_{\rm local} < 
S_{\rm peak} / \sigma_{\rm fit}$) are a consequence of the same problem. \\
In mosaic regions where local noise is significantly larger, 
we applied a 
$6\sigma_{\rm local}$ cut-off if $\sigma_{\rm local}/\sigma_{\rm fit}
\geq 1.2$ (assuming a normal distribution for 
the noise, the probability to get a local to average noise value $\geq 1.2$ 
is $\leq 0.1\%$). 
This resulted in the rejection of 32 sources (see the region in 
Fig.~\ref{fig-lrms} defined by
$S_{\rm peak}\geq 6\sigma_{\rm fit}$, $S_{\rm peak} < 6\sigma_{\rm local}$
and $\sigma_{\rm local}/\sigma_{\rm fit} \geq 1.2$). \\ 
The criteria discussed above proved to be very effective in
selecting out noise artifacts from the catalogue. Nevertheless a few (6) 
sources, which satisfy both the average and the 
local noise constraints are, however, evident noise artifacts at visual 
inspection.
Such objects have been rejected from the final catalogue.  \\
The adopted criteria for the final catalogue definition also 
allowed us to significantly reduce the number of poor 
Gaussian fits (see Sect.~\ref{sec-refit}) since a large fraction of them 
($\sim 65\%$) have fainter $S_{\rm peak}$: we are left with
50 poor Gaussian fits (flagged {\it `S*'}, see Sect.~\ref{sec-cformat}) 
in the final catalogue.\\
A total number of 3172 source components entered the final catalogue.
Some of them have to be considered different components of 
a unique source and, as discussed later in Sect.~\ref{sec-mult}, the 
number of distinct sources in the ATESP catalogue is 2960.

\subsection{Deconvolution}\label{sec-deconv}

\begin{figure}[t]
\resizebox{\hsize}{!}{\includegraphics{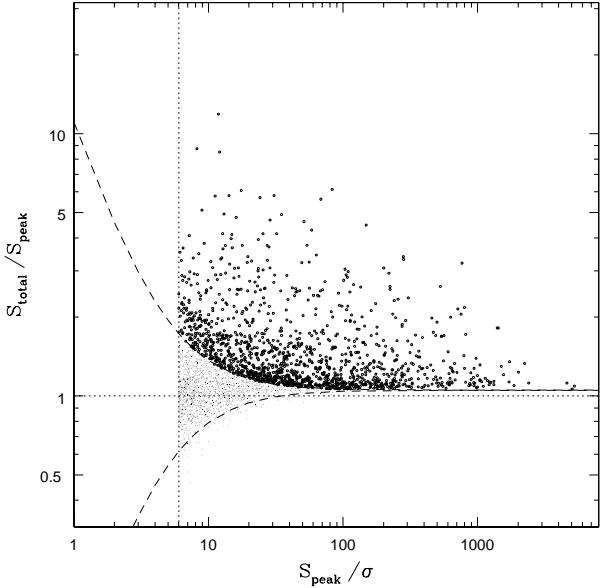}}
\caption{Ratio of the integrated flux $S_{\rm total}$ to the peak one 
$S_{\rm peak}$ as a function of the source signal-to-noise. Dotted lines
indicate the $6\sigma_{\rm fit}$ cut-off adopted for the catalogue definition
(vertical line) and the $S_{\rm total}=S_{\rm peak}$ locus (horizontal 
line) respectively. Also shown are the lower and upper envelopes 
(dashed lines)of the 
flux ratio distribution containing $\sim 90\%$ of the 
unresolved sources (dots). Open circles indicate extended sources.
\label{fig-stspratio}}
\end{figure}

The ratio of the integrated flux to the peak flux is a direct measure of the 
extension of a radio source: 
\begin{equation}
S_{\rm total}/S_{\rm peak}=\theta_{\rm min} \; \theta_{\rm maj} /b_{\rm min} \; b_{\rm maj} 
\end{equation}
where $\theta_{\rm min}$ and $\theta_{\rm maj}$ are the source FWHM axes and $b_{\rm min}$ 
and $b_{\rm maj}$ are the synthesized beam FWHM axes.
The flux ratio can therefore be used to discriminate between extended (larger 
than the beam) and point-like sources. \\
In Fig.~\ref{fig-stspratio} we have plotted the flux 
ratio as a function of the signal-to-noise for all the sources
(or source components) in the ATESP catalogue. 
The flux density ratio has a skew distribution, with a tail towards high flux
ratios due to extended sources. Values for 
$S_{\rm total}/S_{\rm peak} < 1$ are due to the influence of the image noise 
on the measure of source sizes (see Sect.~\ref{sec-errors}). 
To establish a criterion 
for extension, such errors have to be taken into account. 
We have determined the lower envelope of the flux ratio distribution (the curve
containing 90\% of the $S_{\rm total}<S_{\rm peak}$ sources) and we have 
mirrored it on the $S_{\rm total}>S_{\rm peak}$ side (upper envelope in 
Fig.~\ref{fig-stspratio}). We have considered as unresolved all 
sources
laying below the upper envelope. The upper envelope can be characterized
by the equation:
\begin{equation}
S_{\rm total}/S_{\rm peak} = 1.05 + 
\left[ \frac{ 10 }{ (S_{\rm peak}/\sigma_{\rm fit})^{1.5}}\right]  \; .
\end{equation}
From this analysis we found that 1864 of the 3172 sources 
(or source components) in the catalogue (i.e. $\sim 60\%$) have to be 
considered unresolved. \\
It is worth noting that the envelope does not converge to 1 going to large 
signal-to-noise values. This is due to the radial smearing effect. It 
systematically reduces the source peak fluxes, yielding larger 
$S_{\rm total}/S_{\rm peak}$ ratios (see discussion in Sect.~\ref{sec-corr}). 
From 
Fig.~\ref{fig-stspratio} we can quantify the smearing effect in $\sim 5\%$
on average. \\
Deconvolved angular sizes are given in the catalogue only for sources above the
upper curve (filled circles in Fig.~\ref{fig-stspratio}). For unresolved 
sources 
(dots in Fig.~\ref{fig-stspratio}) deconvolved angular sizes are set to zero.\\
Note that no bandwidth correction to deconvolved sizes has been applied.
Correcting for such effect would be somewhat complicated by the fact that 
each source in the radio mosaics is a sum of contributions from several 
single pointings.

\subsection{Multiple Sources}\label{sec-mult}

\begin{figure}[t]
\resizebox{\hsize}{!}{\includegraphics{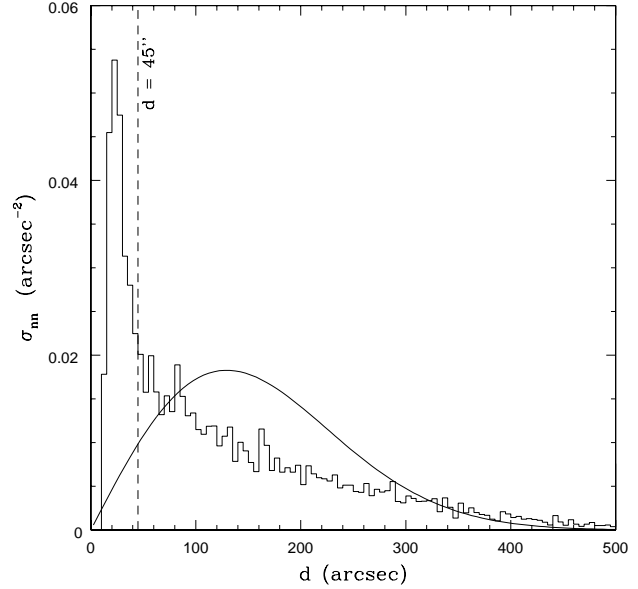}}
\caption[]{Nearest neighbor pair density distribution as a function of distance
(arcsec) for the ATESP radio sources (histogram) and the expected one assuming
a random distribution of the sources in the sky (solid line). The expected
distribution has been scaled so as to have the same area below the curve and 
the observed histogram. The excess
at small distances is due to physical associations, and is compensated by a
deficiency at intermediate distances (at $d\simeq 80-300\arcsec$). 
Edge effects explain the discrepancy between the expected distribution and 
the observed one at very large distances ($d>400\arcsec$).
The vertical 
dashed line indicates the $d=45\arcsec$ cut applied to discriminate between 
real and unreal physical associations.
\label{fig-doubles}}
\end{figure}

\begin{figure}[t]
\resizebox{\hsize}{!}{\includegraphics{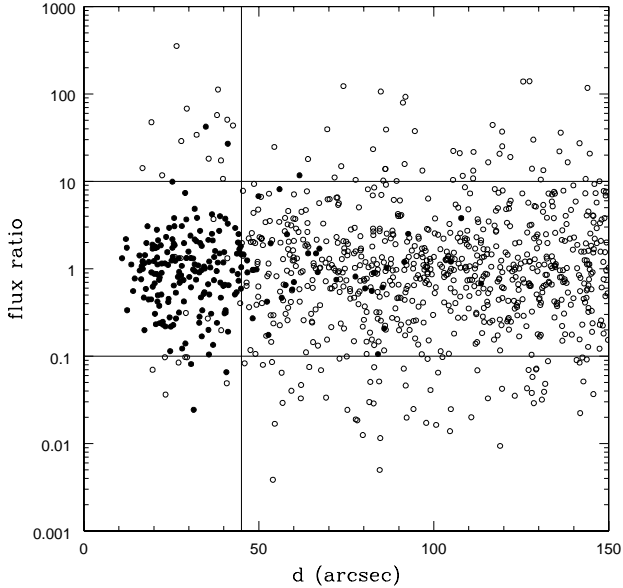}}
\caption[]{Pair components' flux ratios as a function of the distance between the
two components for all the $d\leq 150\arcsec$ pairs in the 
catalogue. Filled circles indicate solely the pairs that have been 
considered as part of a unique multiple source in the final form of the 
catalogue (see text for the criteria adopted for multiple sources' 
definition).
\label{fig-fratio}}
\end{figure}

In Fig.~\ref{fig-doubles} the (nearest neighbor) pair 
density distribution is shown as a function
of distance (histogram). Also indicated is the expected distribution if all 
the sample sources 
(components) were randomly distributed in the sky.  
The expected distribution
has been scaled so as to have the same area below the curve and the observed 
histogram. The excess at small 
distances is clearly due to physical associations and, because of the 
normalization chosen, is compensated by a 
deficiency at larger distances (between $80\arcsec$ and $300\arcsec$). \\
All the components closer than $45\arcsec$ (\emph{i.e.} about three times the 
beam size) have been considered as possibly belonging to a unique double 
source. Triple sources are defined whenever one additional component is closer
than $45\arcsec$ to (at least) one of the pair components. For multiple 
sources the same criterion is applied iteratively. 
Applying this distance constraint we expect that $\sim 20\%$ of the pairs 
are random superpositions. \\
The flux ratio distribution between the pair components has a large spread
at all distances (see Fig.~\ref{fig-fratio}). 
To reduce the contamination we have discarded
all the pairs with flux ratio larger than a factor 10. 
For triples and multiple sources the probable core is not considered when
computing the flux ratios. \\
A few departures from the adopted criteria are present in the catalogue. 
For example the triple 
source ATESP J005620-394145 and the double source ATESP J011029-393253 have
$d\leq 45\arcsec$ but do not satisfy the flux ratio constraint.
All exceptions are 
based on source geometry considerations and/or the analysis of the source 
field. \\
In order to increase the multiple sources' sub-sample completeness, we added 
31 sources with distances $45\arcsec < d < 150\arcsec$, which show 
clear signs of physical associations between their components (see 
Fig.~\ref{fig-examples} for some examples). No flux
ratio constraints have been applied to such sources. 
In Fig.~\ref{fig-fratio}
are shown the flux ratios for all the pairs in the final sample of 
multiple sources (filled circles). \\
As a final result we have 189 multiple sources: 168 doubles, 19 triples
and 2 sources with four components. 
As a consequence, the initial list of 3172 
radio components results in a catalogue of 2960 distinct radio 
sources. 

\begin{table*}[p]
\caption[]{The Radio Catalogue: First Page}
\label{tab-cat}
\begin{flushleft}
\begin{tabular}{lllrrrrrl}
\hline\hline\noalign{\smallskip}
 \multicolumn{1}{c}{IAU Name}   & \multicolumn{1}{c}{R.A.}
& \multicolumn{1}{c}{DEC} & \multicolumn{1}{c}{$S_{\rm peak}$} & \multicolumn{1
}{c}{$S_{\rm total}$} & \multicolumn{1}{c}{$\Theta_{\rm maj}$} & \multicolumn{1}{c}{$\Theta_{\rm min}$} & \multicolumn{1}{c}{P.A.} \\
 \multicolumn{1}{c}{} & \multicolumn{2}{c}{(J2000)}  & \multicolumn{2}{c}{mJy} & \multicolumn{2}{c}{arcsec} & \multicolumn{1}{c}{degr.} \\
\noalign{\smallskip}
\hline\noalign{\smallskip} 
ATESP J223235-402642 & 22:32:35.62 & -40:26:42.1 & 1.26 & 6.07 & 27.13 & 13.55 & 28.9 & S \\
ATESP J223237-393113 & 22:32:37.71 & -39:31:13.2 & 0.69 & 0.71 & 0.00 & 0.00 & 0.0 & S \\
ATESP J223238-394102 & 22:32:38.54 & -39:41:02.9 & 0.48 & 0.58 & 0.00 & 0.00 & 0.0 & S \\
ATESP J223242-393054 & 22:32:42.10 & -39:30:54.3 & 2.33 & 2.46 & 0.00 & 0.00 & 0.0 & S \\
ATESP J223248-401345 & 22:32:48.24 & -40:13:45.9 & 0.95 & 0.83 & 0.00 & 0.00 & 0.0 & S \\
ATESP J223250-394059 & 22:32:50.79 & -39:40:59.7 & 0.57 & 0.93 & 10.47 & 0.00 & -88.4 & S \\
ATESP J223252-401925 & 22:32:52.45 & -40:19:25.1 & 0.66 & 0.46 & 0.00 & 0.00 & 0.0 & S \\
ATESP J223254-393652 & 22:32:54.54 & -39:36:52.0 & 1.03 & 1.70 & 9.89 & 5.39 & 48.3 & S \\
ATESP J223255-395717 & 22:32:55.51 & -39:57:17.3 & 0.51 & 0.45 & 0.00 & 0.00 & 0.0 & S \\
ATESP J223256-402010 & 22:32:56.10 & -40:20:10.3 & 1.13 & 1.08 & 0.00 & 0.00 & 0.0 & S \\
ATESP J223301-393017 & 22:33:01.55 & -39:30:17.1 & 7.83 & 8.67 & 3.39 & 2.66 & -57.0 & S \\
ATESP J223302-402817 & 22:33:02.31 & -40:28:17.5 & 0.66 & 0.71 & 0.00 & 0.00 & 0.0 & S \\
ATESP J223303-401629 & 22:33:03.07 & -40:16:29.2 & 0.99 & 1.27 & 5.30 & 4.74 & 62.9 & S \\
ATESP J223304-395639 & 22:33:04.92 & -39:56:39.0 & 1.85 & 2.94 & 23.51 & -- & -- & M \\
ATESP J223304-395639A & 22:33:04.43 & -39:56:41.4 & 1.85 & 2.18 & 6.66 & 1.43 & -24.2 & S \\
ATESP J223304-395639B & 22:33:06.31 & -39:56:32.2 & 0.80 & 0.76 & 0.00 & 0.00 & 0.0 & S \\
ATESP J223313-400216 & 22:33:13.42 & -40:02:16.1 & 7.60 & 9.18 & 4.94 & 3.65 & -45.7 & S \\
ATESP J223314-394942 & 22:33:14.42 & -39:49:42.8 & 1.28 & 1.33 & 0.00 & 0.00 & 0.0 & S \\
ATESP J223316-393124 & 22:33:16.90 & -39:31:24.4 & 0.52 & 0.68 & 0.00 & 0.00 & 0.0 & S \\
ATESP J223317-393235 & 22:33:17.19 & -39:32:35.0 & 2.40 & 3.37 & 9.29 & 4.22 & 9.4 & S \\
ATESP J223320-394713 & 22:33:20.14 & -39:47:13.8 & 0.96 & 1.08 & 0.00 & 0.00 & 0.0 & S \\
ATESP J223322-401710 & 22:33:22.93 & -40:17:10.7 & 4.92 & 6.26 & 7.56 & 3.34 & 14.7 & S \\
ATESP J223327-395836 & 22:33:27.45 & -39:58:36.9 & 3.24 & 5.16 & 13.60 & 3.39 & -5.9 & S \\
ATESP J223327-394541 & 22:33:27.73 & -39:45:41.7 & 2.74 & 5.70 & 21.21 & -- & -- & M \\
ATESP J223327-394541A & 22:33:27.08 & -39:45:40.2 & 2.74 & 3.65 & 5.98 & 4.83 & 60.7 & S \\
ATESP J223327-394541B & 22:33:28.89 & -39:45:44.4 & 1.29 & 2.05 & 10.04 & 0.00 & -73.8 & S \\
ATESP J223329-402019 & 22:33:29.10 & -40:20:19.6 & 1.41 & 1.82 & 6.85 & 3.75 & -32.1 & S \\
ATESP J223330-395233 & 22:33:30.97 & -39:52:33.4 & 0.50 & 0.53 & 0.00 & 0.00 & 0.0 & S \\
ATESP J223335-401337 & 22:33:35.21 & -40:13:37.7 & 0.63 & 0.71 & 0.00 & 0.00 & 0.0 & S \\
ATESP J223337-394253 & 22:33:37.01 & -39:42:53.8 & 2.58 & 2.79 & 0.00 & 0.00 & 0.0 & S \\
ATESP J223338-392919 & 22:33:38.49 & -39:29:19.6 & 5.26 & 5.62 & 0.00 & 0.00 & 0.0 & S \\
ATESP J223339-393131 & 22:33:39.86 & -39:31:31.0 & 1.37 & 1.32 & 0.00 & 0.00 & 0.0 & S \\
ATESP J223343-393811 & 22:33:43.21 & -39:38:11.2 & 1.00 & 1.05 & 0.00 & 0.00 & 0.0 & S \\
ATESP J223343-402307 & 22:33:43.93 & -40:23:07.4 & 0.90 & 1.10 & 0.00 & 0.00 & 0.0 & S \\
ATESP J223345-402815 & 22:33:45.23 & -40:28:15.6 & 0.56 & 0.65 & 0.00 & 0.00 & 0.0 & S \\
ATESP J223346-393322 & 22:33:46.35 & -39:33:22.9 & 1.26 & 1.23 & 0.00 & 0.00 & 0.0 & S \\
ATESP J223351-394040 & 22:33:51.19 & -39:40:40.9 & 1.12 & 1.17 & 0.00 & 0.00 & 0.0 & S \\
ATESP J223356-401949 & 22:33:56.57 & -40:19:49.7 & 1.71 & 2.28 & 36.37 & -- & -- & M \\
ATESP J223356-401949A & 22:33:55.33 & -40:20:11.8 & 0.52 & 0.63 & 0.00 & 0.00 & 0.0 & S \\
ATESP J223356-401949B & 22:33:57.05 & -40:19:41.2 & 1.71 & 1.65 & 0.00 & 0.00 & 0.0 & S \\
ATESP J223358-400642 & 22:33:58.77 & -40:06:42.4 & 1.29 & 1.20 & 0.00 & 0.00 & 0.0 & S \\
ATESP J223401-402310 & 22:34:01.25 & -40:23:10.8 & 0.58 & 1.18 & 10.98 & 9.14 & 32.8 & S \\
ATESP J223401-393448 & 22:34:01.87 & -39:34:48.9 & 1.69 & 2.09 & 5.28 & 3.10 & -72.8 & S \\
ATESP J223402-402357 & 22:34:02.36 & -40:23:57.7 & 0.59 & 0.57 & 0.00 & 0.00 & 0.0 & S \\
ATESP J223402-400017 & 22:34:02.64 & -40:00:17.3 & 1.70 & 1.99 & 5.37 & 2.86 & -28.2 & S \\
ATESP J223404-395831 & 22:34:04.00 & -39:58:31.5 & 6.16 & 7.45 & 5.53 & 3.43 & 29.7 & S \\
ATESP J223404-393358 & 22:34:04.08 & -39:33:58.9 & 0.85 & 0.90 & 0.00 & 0.00 & 0.0 & S \\
ATESP J223407-393721 & 22:34:07.34 & -39:37:21.9 & 3.14 & 6.41 & 25.71 & -- & -- & M \\
ATESP J223407-393721A & 22:34:06.35 & -39:37:27.4 & 2.03 & 3.23 & 8.90 & 6.08 & 38.9 & S \\
ATESP J223407-393721B & 22:34:08.36 & -39:37:16.3 & 3.14 & 3.19 & 0.00 & 0.00 & 0.0 & S \\
ATESP J223409-394258 & 22:34:09.58 & -39:42:58.2 & 0.71 & 1.11 & 10.16 & 5.23 & -15.4 & S \\
ATESP J223410-394427 & 22:34:10.82 & -39:44:27.6 & 1.54 & 1.59 & 0.00 & 0.00 & 0.0 & S \\
ATESP J223412-400254 & 22:34:12.80 & -40:02:54.6 & 0.66 & 0.69 & 0.00 & 0.00 & 0.0 & S \\
ATESP J223413-393242 & 22:34:13.28 & -39:32:42.4 & 1.58 & 1.88 & 6.21 & 0.00 & 59.9 & S \\
ATESP J223413-393650 & 22:34:13.74 & -39:36:50.2 & 0.48 & 1.71 & 25.27 & 8.28 & 38.4 & S \\
ATESP J223413-395651 & 22:34:13.80 & -39:56:51.7 & 0.52 & 0.56 & 0.00 & 0.00 & 0.0 & S \\
ATESP J223420-393150 & 22:34:20.47 & -39:31:50.5 & 0.55 & 0.73 & 0.00 & 0.00 & 0.0 & S \\
\noalign{\smallskip}
\hline\hline
\end{tabular}
\end{flushleft}
\end{table*}

\subsection{Non-Gaussian Sources}\label{sec-complex}

In the final catalogue we have 23 non-Gaussian sources  
whose parameters have been  defined as discussed in Sect.~\ref{sec-refit}. 
In particular we notice that 
positions refer to peak positions,  
which, for non-Gaussian sources does not
necessarily correspond to the position of the core. We also notice that we can
have non-Gaussian components in multiple sources. Some examples of single 
and multiple non Gaussian sources are shown in Fig.~\ref{fig-examples}. 

\begin{figure*}[t]
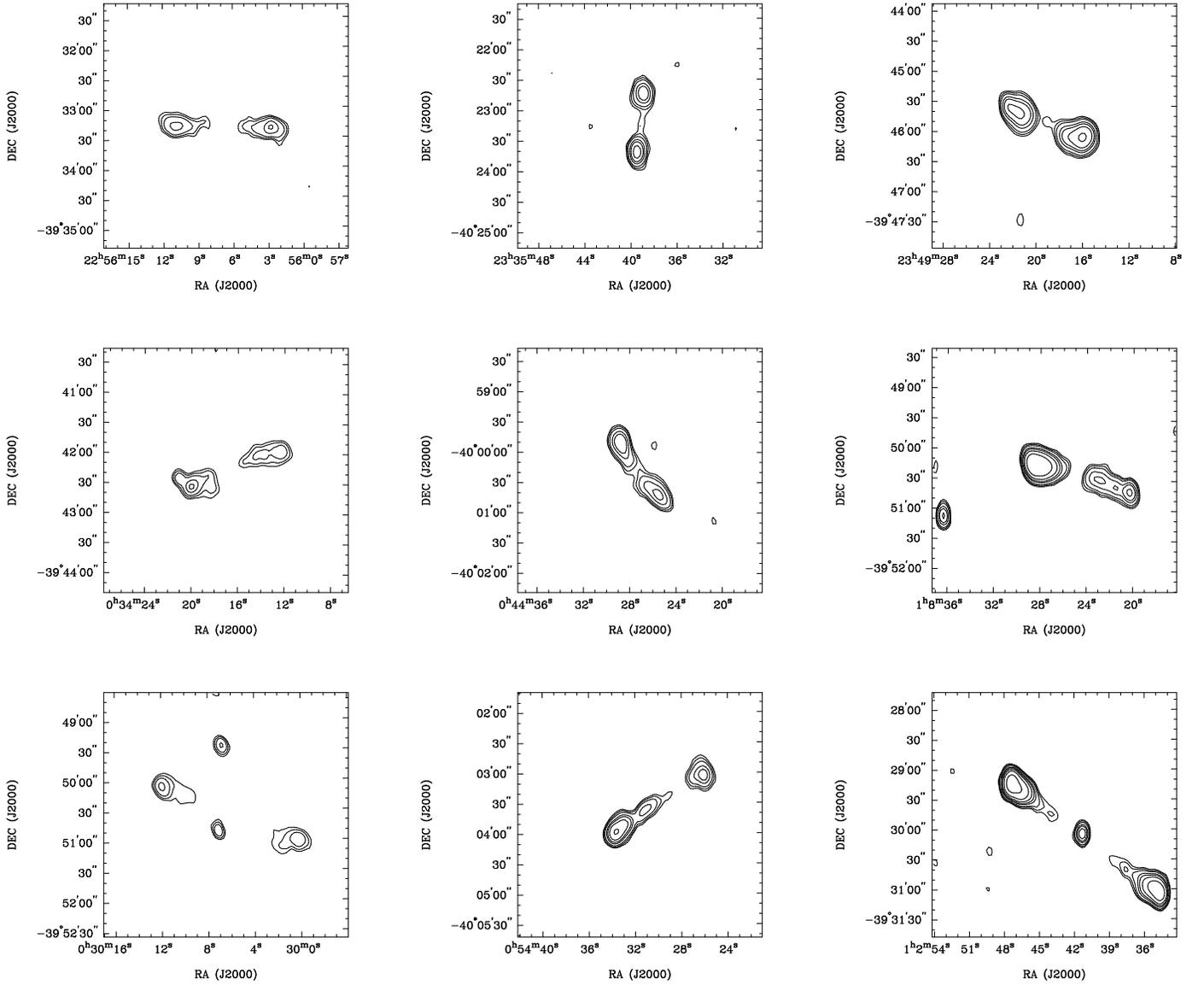

\resizebox{4.5cm}{!}{\includegraphics{H2076f7a.ps}}
\resizebox{4.5cm}{!}{\includegraphics{H2076f7b.ps}}
\resizebox{4.5cm}{!}{\includegraphics{H2076f7c.ps}}
\resizebox{4.5cm}{!}{\includegraphics{H2076f7d.ps}}
\resizebox{4.5cm}{!}{\includegraphics{H2076f7e.ps}}
\resizebox{4.5cm}{!}{\includegraphics{H2076f7f.ps}}
\resizebox{4.5cm}{!}{\includegraphics{H2076f7g.ps}}
\resizebox{4.5cm}{!}{\includegraphics{H2076f7h.ps}}
\resizebox{4.5cm}{!}{\includegraphics{H2076f7i.ps}}
\hfill
\parbox[b]{18cm}{
\caption{Examples of very large 
($d>45\arcsec$, see text) multi-component ATESP sources  
and of non-Gaussian ATESP sources. For direct comparison $4\arcmin \times 
4\arcmin$ contour images are presented for all sources. For each source, the 
contour levels are at 3, 6, 10, 20, 30, 50, 100, 300 times the average 
rms flux density of the mosaic where the source has been detected.
\emph{1$^{st}$ and 2$^{nd}$ row:} double sources. 
\emph{$3^{rd}$ row:} triple sources.}  
\label{fig-examples}}
\end{figure*}
\begin{figure*}[t]
\addtocounter{figure}{-1}
\resizebox{4.5cm}{!}{\includegraphics{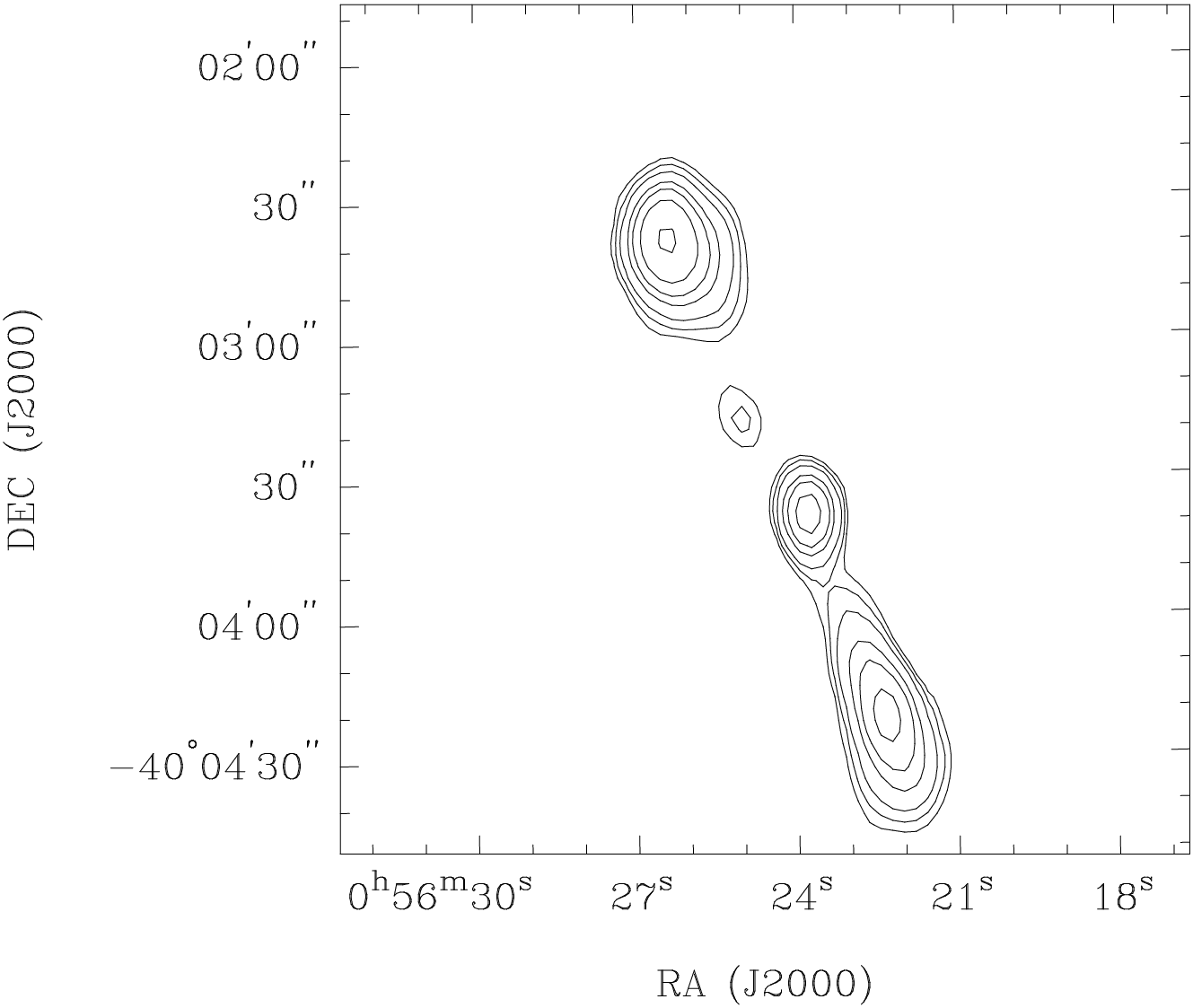}}
\resizebox{4.5cm}{!}{\includegraphics{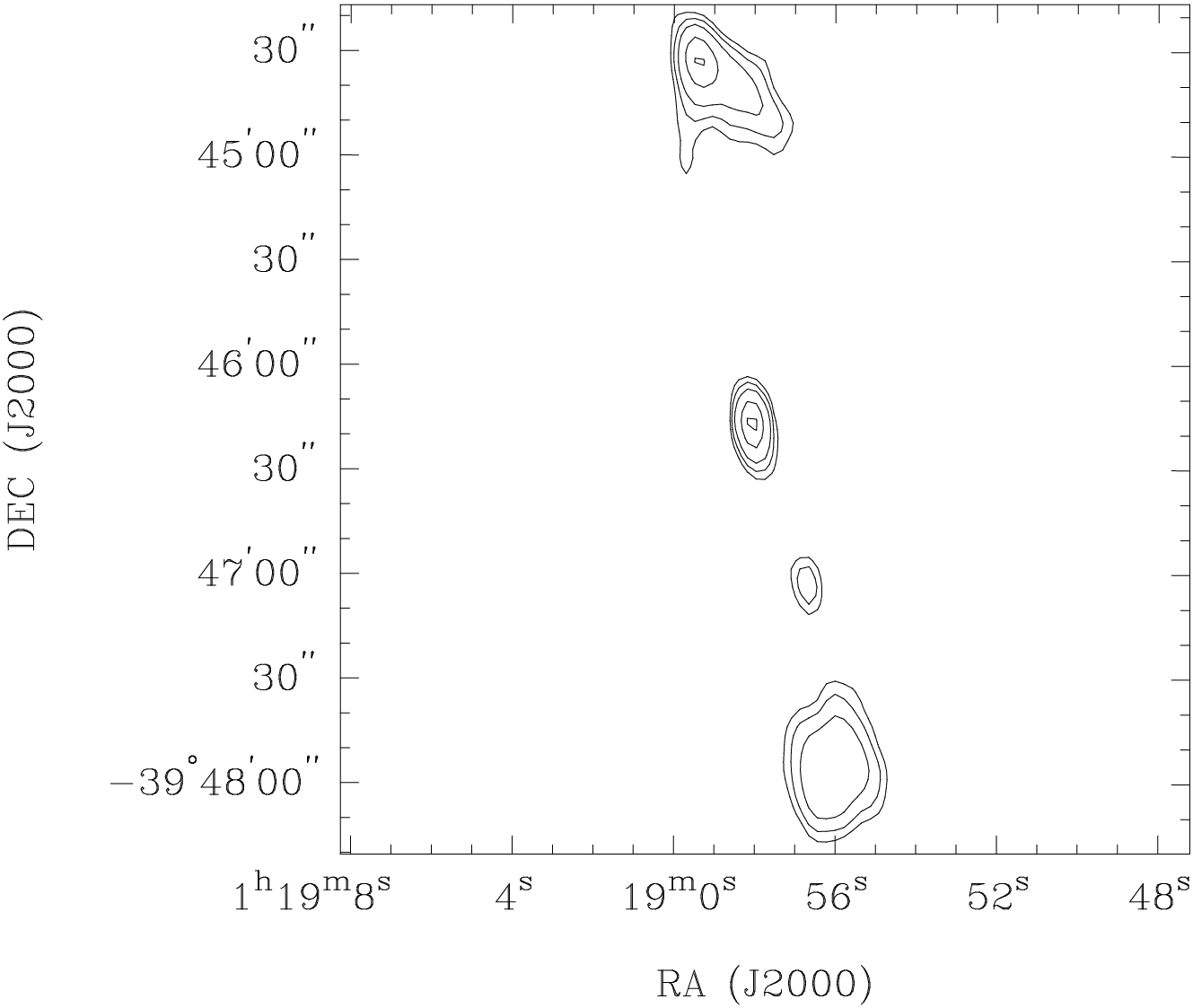}}
\resizebox{4.5cm}{!}{\includegraphics{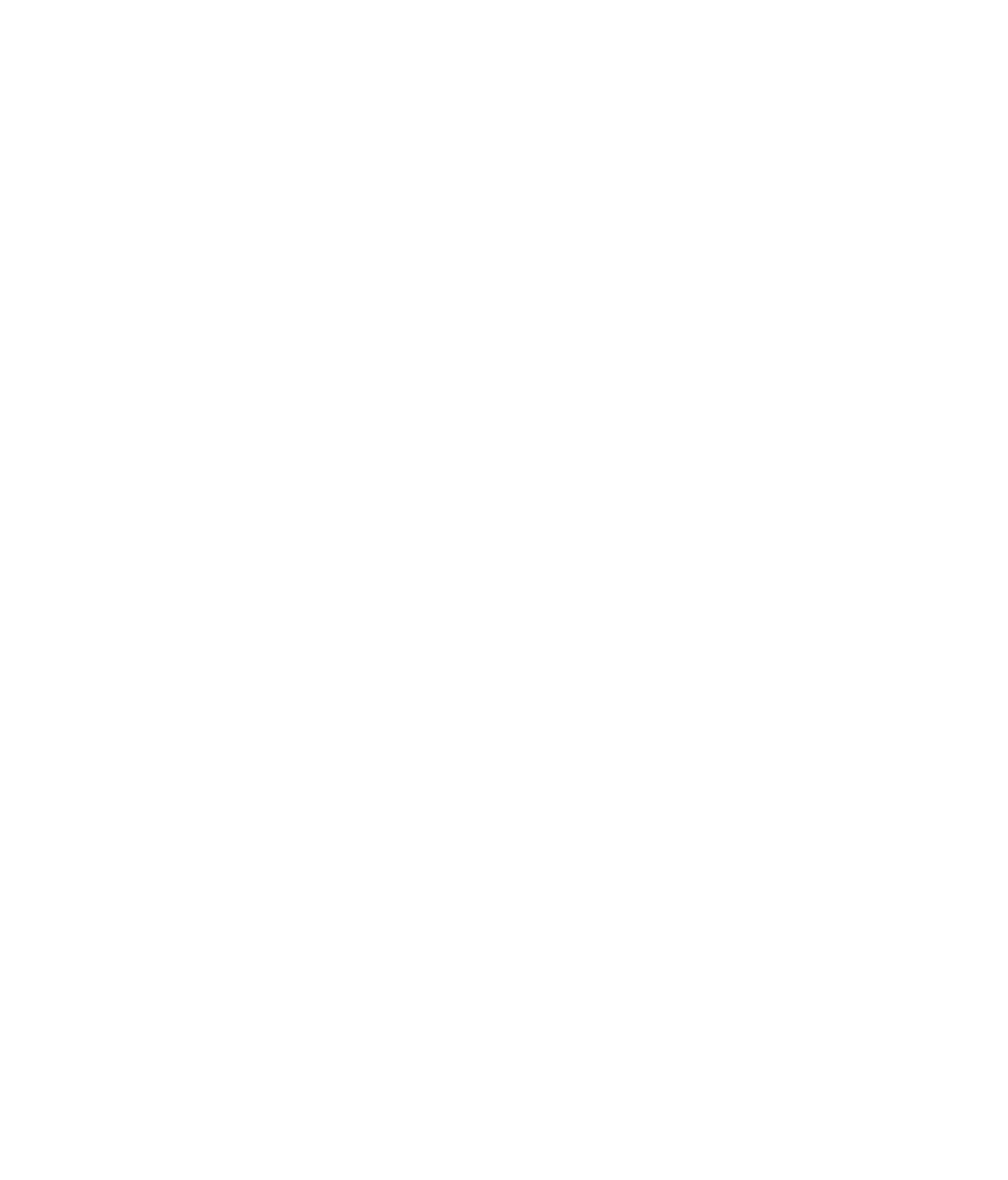}}
\resizebox{4.5cm}{!}{\includegraphics{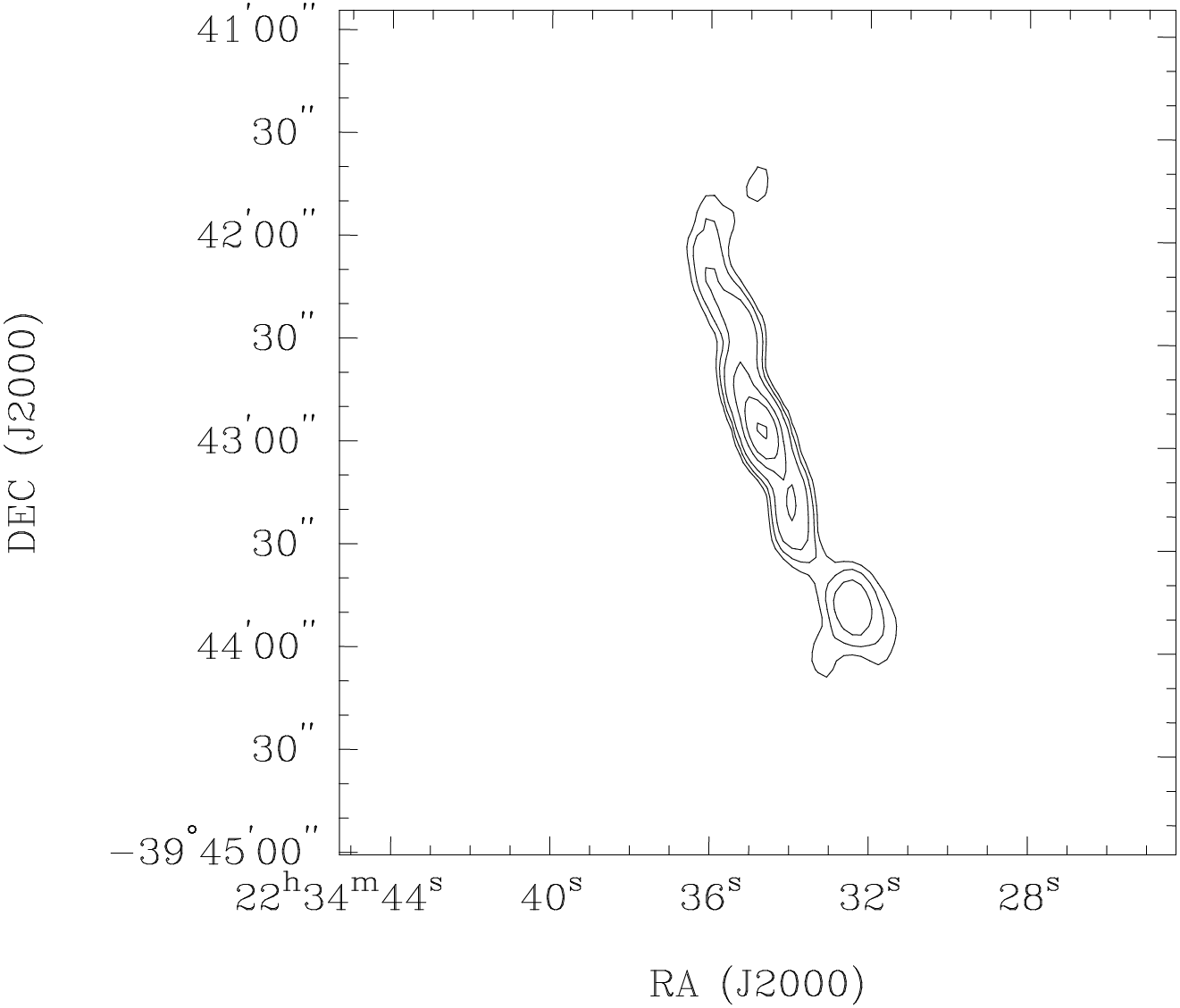}}
\resizebox{4.5cm}{!}{\includegraphics{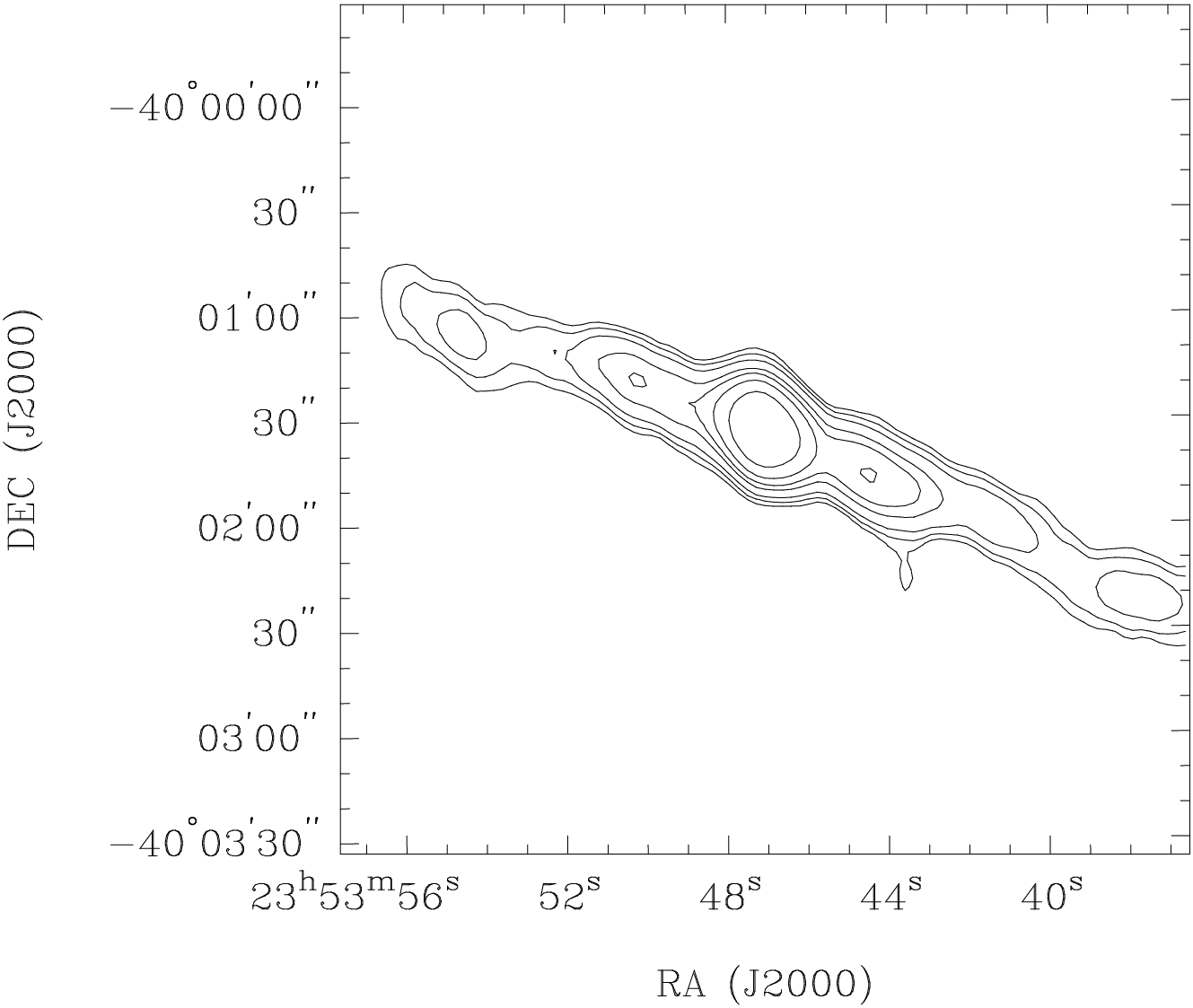}}
\resizebox{4.5cm}{!}{\includegraphics{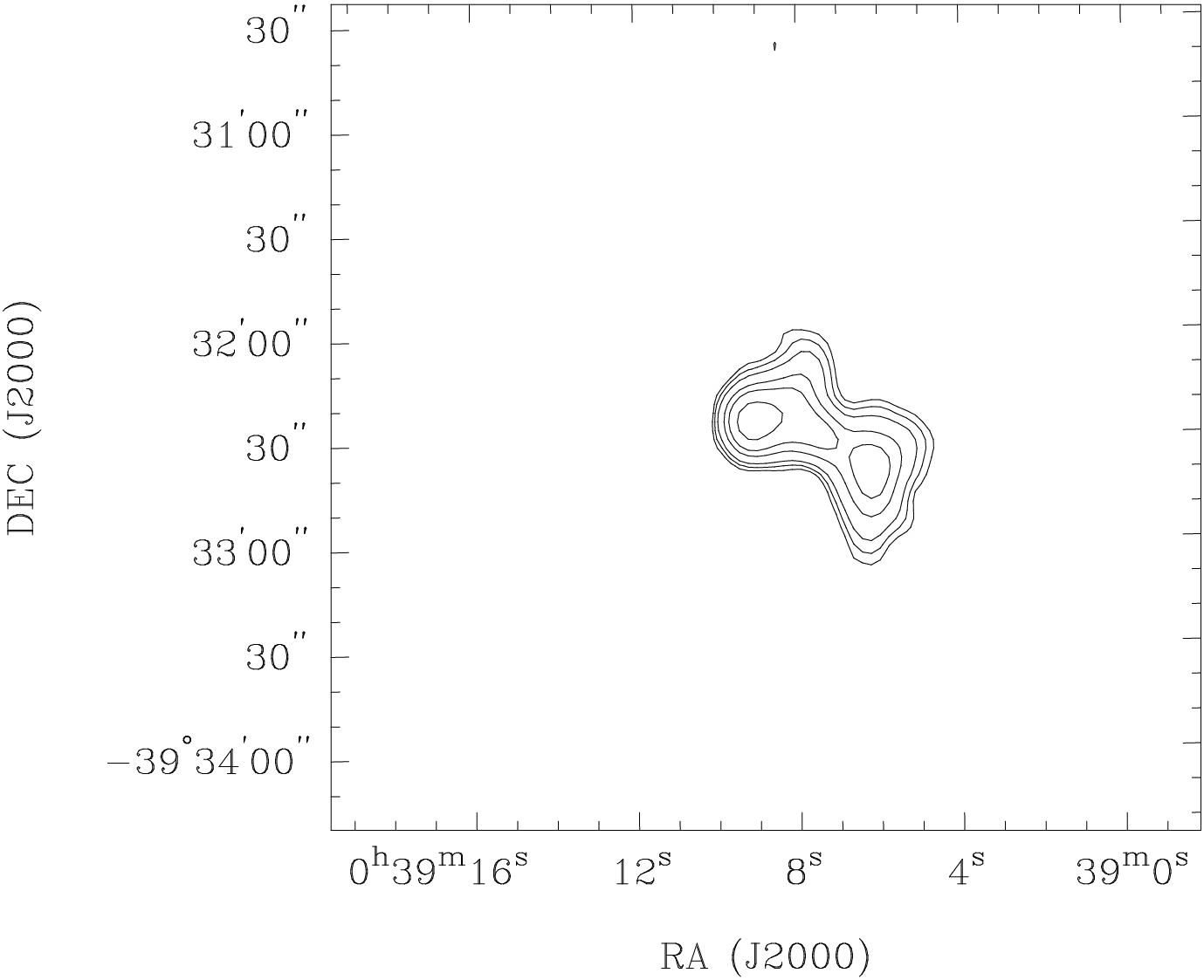}}
\resizebox{4.5cm}{!}{\includegraphics{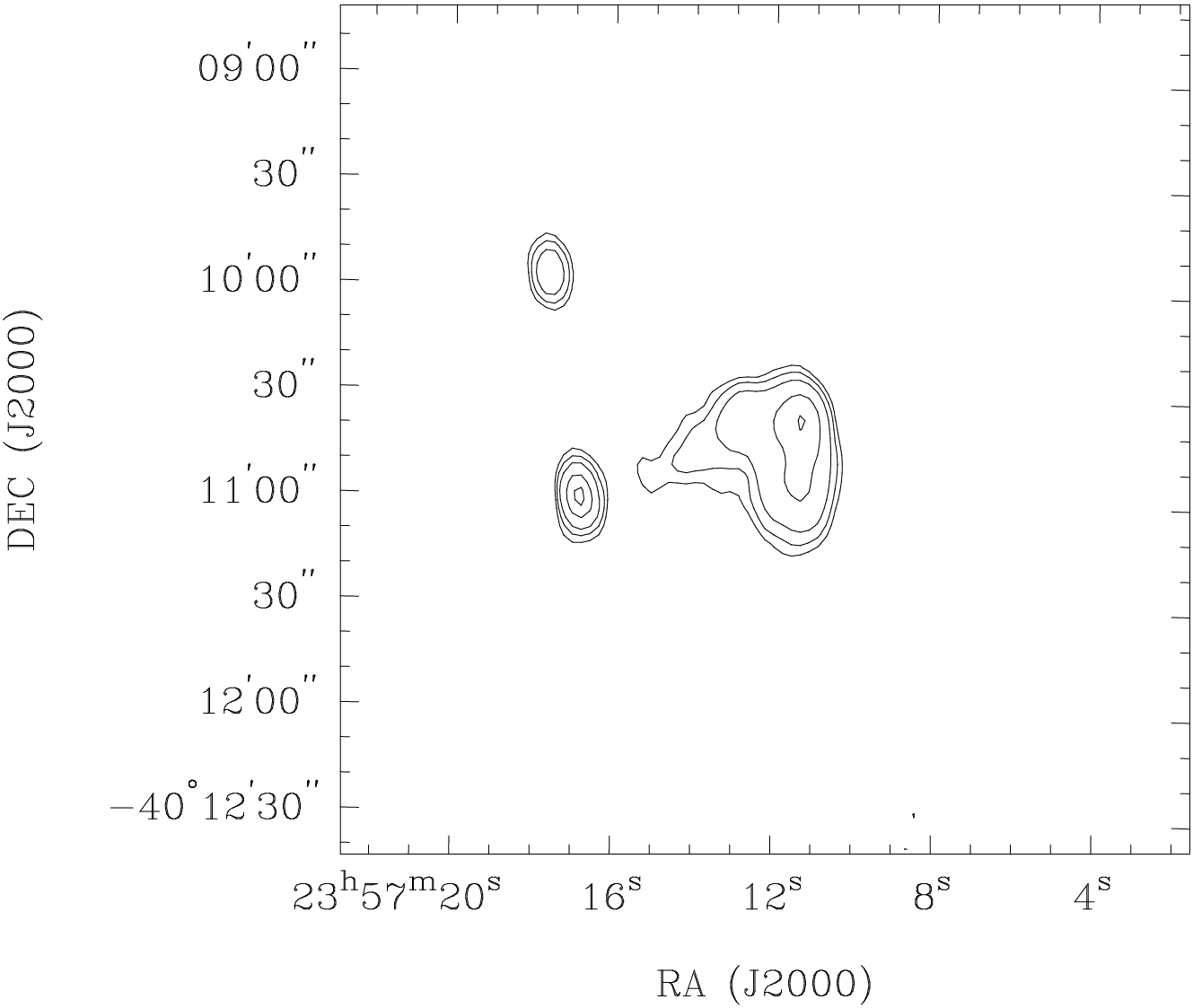}}
\resizebox{4.5cm}{!}{\includegraphics{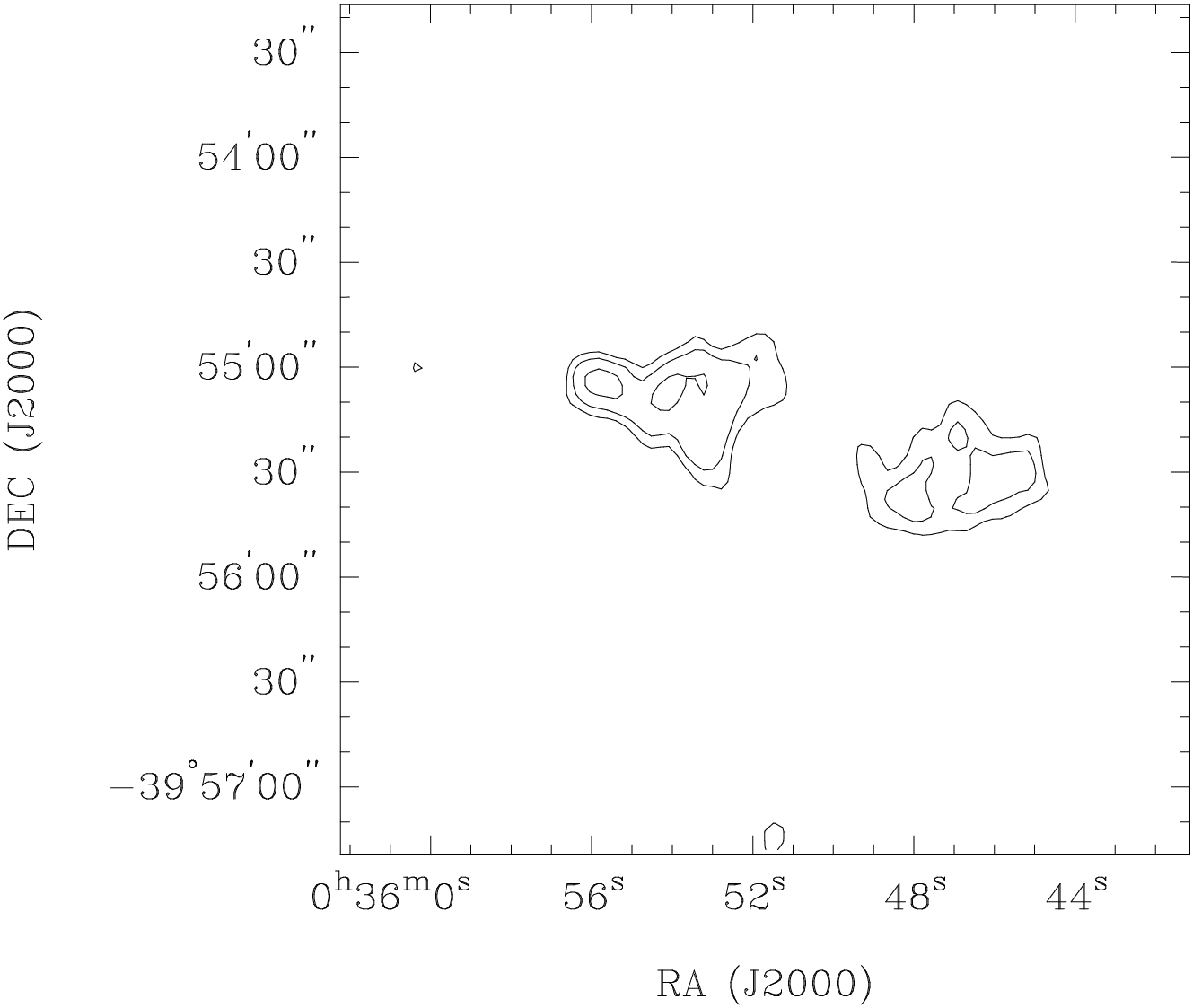}}
\resizebox{4.5cm}{!}{\includegraphics{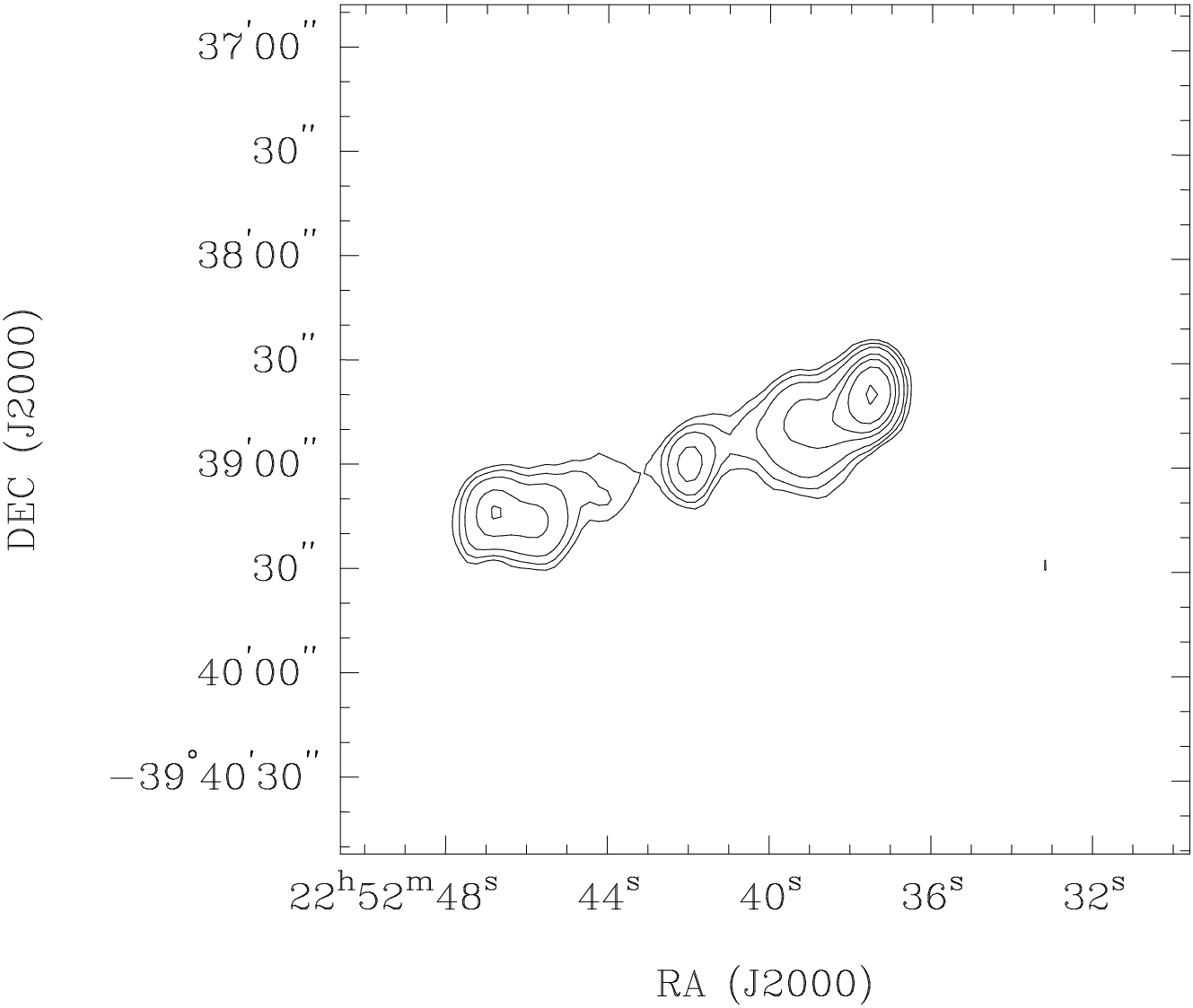}}
\hfill
\parbox[b]{18cm}{
\caption{{\bf Continued.} Examples of very large 
($d>45\arcsec$, see text) multi-component ATESP sources  
and of non-Gaussian ATESP sources. For direct comparison $4\arcmin \times 
4\arcmin$ contour images are presented for all sources. For each source, the 
contour levels are at 3, 6, 10, 20, 30, 50, 100, 300 times the average 
rms flux density of the mosaic where the source has been detected.
\emph{1$^{st}$ row:} the two 4-component sources in the 
ATESP catalogue. \emph{$2^{nd}$ row:} single-component non Gaussian sources.
\emph{$3^{rd}$ row:} double sources with one (or two) non-Gaussian 
components.}} 
\end{figure*}

\subsection{The Catalogue Format}\label{sec-cformat}

The electronic version of the 
full radio catalog is available through the ATESP page at 
{\tt http://www.ira.bo.cnr.it}. 
Its first page is shown as an example in Table~\ref{tab-cat}. 
The source catalogue is sorted on right ascension.
The format is the following:

\indent {\it Column (1) -} Source IAU name. Different  
components of multiple sources are labeled `A', `B', etc. \\ 
\indent {\it Column (2) and (3) -} Source position: Right Ascension and 
Declination (J2000). \\
\indent {\it Column (4) and (5) -} Source peak ($S_{\rm peak}$) and 
integrated ($S_{\rm total}$) flux densities in mJy (Baars
et al. \cite{Baarsetal77} scale). The flux densities are not corrected 
for the systematic effects discussed in Sect.~\ref{sec-corr}.\\
\indent {\it Column (6) and (7) -} Intrinsic (deconvolved from the beam) 
source angular size. Full width half maximum
of the major ($\Theta_{\rm maj}$) and minor ($\Theta_{\rm min}$) axes in 
arcsec. 
Zero values 
refer to unresolved sources (see Sect.~\ref{sec-deconv} for more
details).\\
\indent {\it Column (8) -} Source position angle (P.A., measured N through E) 
for the major axis 
in degrees. \\
\indent {\it Column (9) -} Flag indicating the fitting procedure and 
parameterization adopted for the source or source component (see 
Sects.~\ref{sec-refit} and \ref{sec-mult}).
{\it `S'} refers to Gaussian fits. {\it `S*'} refers to poor Gaussian fits. 
{\it `E'} refers to non-Gaussian sources. {\it `M'} refers to multiple 
sources (see below). 

\noindent
The parameters listed for non-Gaussian sources  
are defined as discussed in Sect.~\ref{sec-refit}. \\
For multiple sources we list all the components (labeled `A', `B', etc.)
preceded by a line (flagged {\it `M'}) giving the position of the radio 
centroid, total flux density and overall angular
size of the source. Source positions have been defined 
as the flux-weighted average position of all the components (source 
centroid). For sources with more than two components the centroid position
has been replaced with the core position whenever the core is clearly 
recognizable. \\
Integrated total source flux densities are computed by summing 
all the component integrated fluxes. \\
The total source angular size is defined as 
\emph{las} (see Sect.~\ref{sec-refit}) and it 
is computed as the maximum distance between the source components. 

\section{Errors in the Source Parameters}\label{sec-errors}

Parameter uncertainties are the quadratic sum of two 
independent terms: the calibration errors, which dominate at high 
signal-to-noise ratios, and the internal errors, due to the presence of noise 
in the maps. The latter dominate at low signal-to-noise ratios. \\
In the following sections we discuss the parameter internal accuracy of our 
source catalogue. Master equations  for total rms error derivation, with 
estimates of the calibration terms are reported in Appendix~\ref{sec-appa}. 

\subsection{Internal accuracy}\label{sec-interr}

In order to quantify the internal errors 
we produced a one square degree residual map by removing all the sources 
detected above $5\sigma$ in the radio mosaic fld20to25. We performed a 
set of Montecarlo simulations by injecting Gaussian sources in the 
residual map at 
random positions and re-extracting them using the same detection algorithm 
used for the survey (IMSAD). 
The Montecarlo simulations were performed by injecting
samples of 30 sources at fixed flux and intrinsic angular size. 
We sampled peak fluxes between $5\sigma$ and $50\sigma$ and 
intrinsic angular sizes 
(FWHM major axis) between $4\arcsec$ and $20\arcsec$.  
Intrinsic sizes were convolved with the synthesized beam 
($8.5\arcsec\times 16.8\arcsec$ for mosaic fld20to25) before injecting the 
source in the residual map. \\
The comparison between the input parameters 
and the ones provided by IMSAD permitted an estimate of the internal 
accuracy of the parameters as a function of source flux and intrinsic 
angular size. 
In particular we could test the accuracy of flux densities, positions and 
angular sizes and estimate 
both the efficiency and the accuracy of the deconvolution algorithm. 

\subsubsection{Flux Densities and Source Sizes}\label{sec-fluxerr}

\begin{figure}[t]
\resizebox{\hsize}{!}{\includegraphics{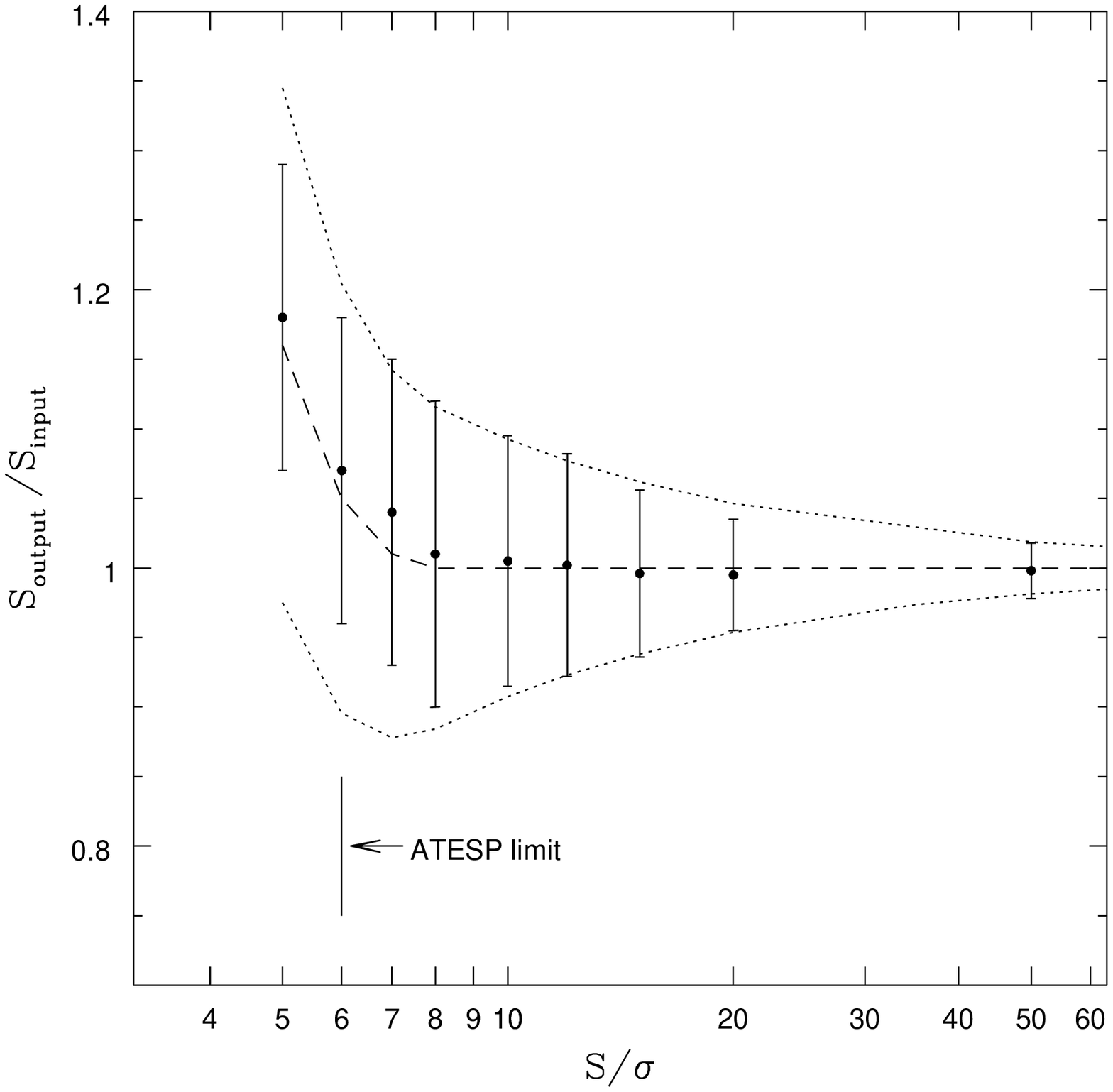}}
\hfill
\caption{Peak flux density internal accuracy for point sources. 
Mean and standard deviation for the output/input 
(IMSAD/injected) ratio, as a function of flux. Expected 
values (see text) are also plotted for both the mean (dashed line) and the rms 
(dotted lines). The flux density cut-off chosen for the ATESP catalogue is
indicated by the vertical solid line.
\label{fig-spsim}}
\end{figure}

\begin{figure}[t]
\resizebox{\hsize}{!}{\includegraphics{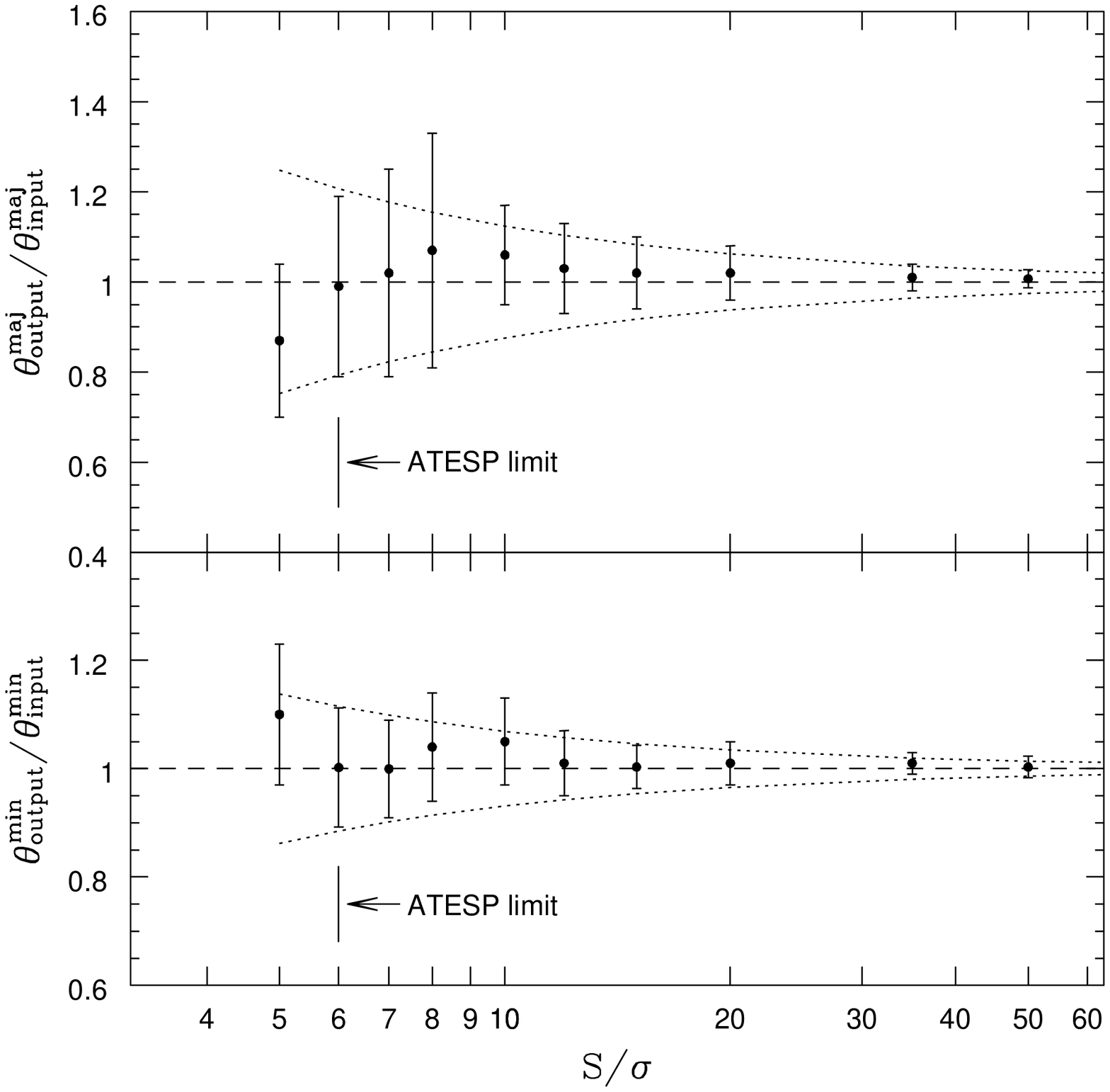}}
\hfill
\caption{Fitted FWHM axes internal accuracy for point sources. Top panel: 
major axis. Bottom panel: minor axis.
Mean and standard deviation for the output/input 
(IMSAD/injected) ratio, as a function of flux. Expected 
values (see text) are also plotted for both the mean (dashed line) and the rms 
(dotted lines). The flux density cut-off chosen for the ATESP catalogue is
indicated by the vertical solid line.
\label{fig-majminsim}}
\end{figure}

The flux density and fitted angular size errors for point sources are shown in 
in Figs.~\ref{fig-spsim}~and~\ref{fig-majminsim} where
we plot the ratio of the parameter value found by IMSAD (output) over the 
injected one (input), as a function of the signal-to-noise ratio. \\
We notice that mean values very far from 1 could indicate the presence of 
systematic effects in the parameter measure. The presence of such systematic
effects is clearly present for peak flux densities in the faintest bins
(see Fig.~\ref{fig-spsim}). 
This is the expected effect of the noise on the catalogue completeness at 
the extraction threshold. 
Due to its Gaussian distribution whenever an injected source falls on a 
noise dip, either the source flux is underestimated or the source goes 
undetected. This 
produces an incompleteness in the faintest bins. As a consequence, the 
measured fluxes are biased toward higher values 
in the incomplete bins, beacause only sources that fall on noise peaks 
have been detected and measured.
We notice that the mean values 
found for $S_{\rm output}/S_{\rm input}$ are in good agreement with the ones 
expected taking into account such an effect (see dashed line). 
It is worth pointing out that our catalogue is only slightly affected by this 
effect because the detection threshold ($4.5\sigma$) is much lower than the 
$6\sigma$-threshold chosen for the catalogue (indicated by the vertical 
solid line in Fig.~\ref{fig-spsim}): at $S_{\rm peak}\geq 6\sigma$ we expect flux 
over-estimations $\leq 5\%$. \\ 
Some systematic effects appear to be present also for the source size at 
$5\sigma$: the major and minor axes tend to be respectively under-  and 
over-estimated (see Fig.~\ref{fig-majminsim}). 
Such effects disappear at $S_{\rm peak} \geq 6\sigma$ (ATESP 
cut-off).
For both the flux densities and the source axes, the rms values measured 
are in very good 
agreement with the ones proposed by Condon (\cite{Condon97}) for elliptical 
Gaussian fitting procedures (for details see Appendix~\ref{sec-appa}):
\begin{eqnarray} 
\sigma(S_{\rm peak})/S_{\rm peak}& = & 0.93 \; \left(\frac{S_{\rm peak}}{\sigma}\right)^{-1} \label{eq-sperr}\\
\sigma(\theta_{\rm maj})/\theta_{\rm maj} & = & 1.24\; \left(\frac{S_{\rm peak}}{\sigma}\right)^{-1} \label{eq-majerr}\\
\sigma(\theta_{\rm min})/\theta_{\rm min} & = & 0.69\;\left(\frac{S_{\rm peak}}{\sigma}\right)^{-1} \label{eq-minerr}
\end{eqnarray}
where we have applied Eqs. (21) and (41) in Condon (\cite{Condon97}) to the 
case of
ATESP point sources (see dotted lines in Figs.~\ref{fig-spsim} and 
\ref{fig-majminsim}). \\
The fact that a source is extended does not affect the 
internal 
accuracy of the fitting algorithm for both the peak flux density and the 
source axes. In other words the errors quoted 
for point sources apply to extended sources as well. \\
However, this is not true for the deconvolution algorithm. The errors for the 
deconvolved source axes depend on both the source flux and intrinsic angular
size.
The higher the flux and the larger the source, the smaller the error. 
In particular, at 1 mJy ($\sim 12\sigma$) the errors are in the range 
$35\%$ -- $10\%$ for angular sizes in the range $6\arcsec$ -- $20\arcsec$. 
For fluxes $>50\sigma$ the errors are always $<10\%$. 
Deconvolved angular sizes are unreliable for very faint sources 
($5-6\sigma$), where only a very small fraction of sources can be 
deconvolved. The deconvolution efficiency increases with the
source flux. In particular, the fraction of deconvolved sources with 
intrinsic dimension $\leq 4\arcsec$ never reaches $100\%$: it goes from 
$3\%$ at the lowest fluxes, to $15\%$ at 1 mJy, to $50\%$ at the highest 
fluxes. We therefore can assume that $4\arcsec$ is
a critical value for deconvolution at the ATESP resolution, 
and that ATESP sources with intrinsic sizes $\leq 4\arcsec$  are to be
considered unresolved. 

\subsubsection{Source Positions}\label{sec-poserr}

\begin{figure}[t]
\resizebox{\hsize}{!}{\includegraphics{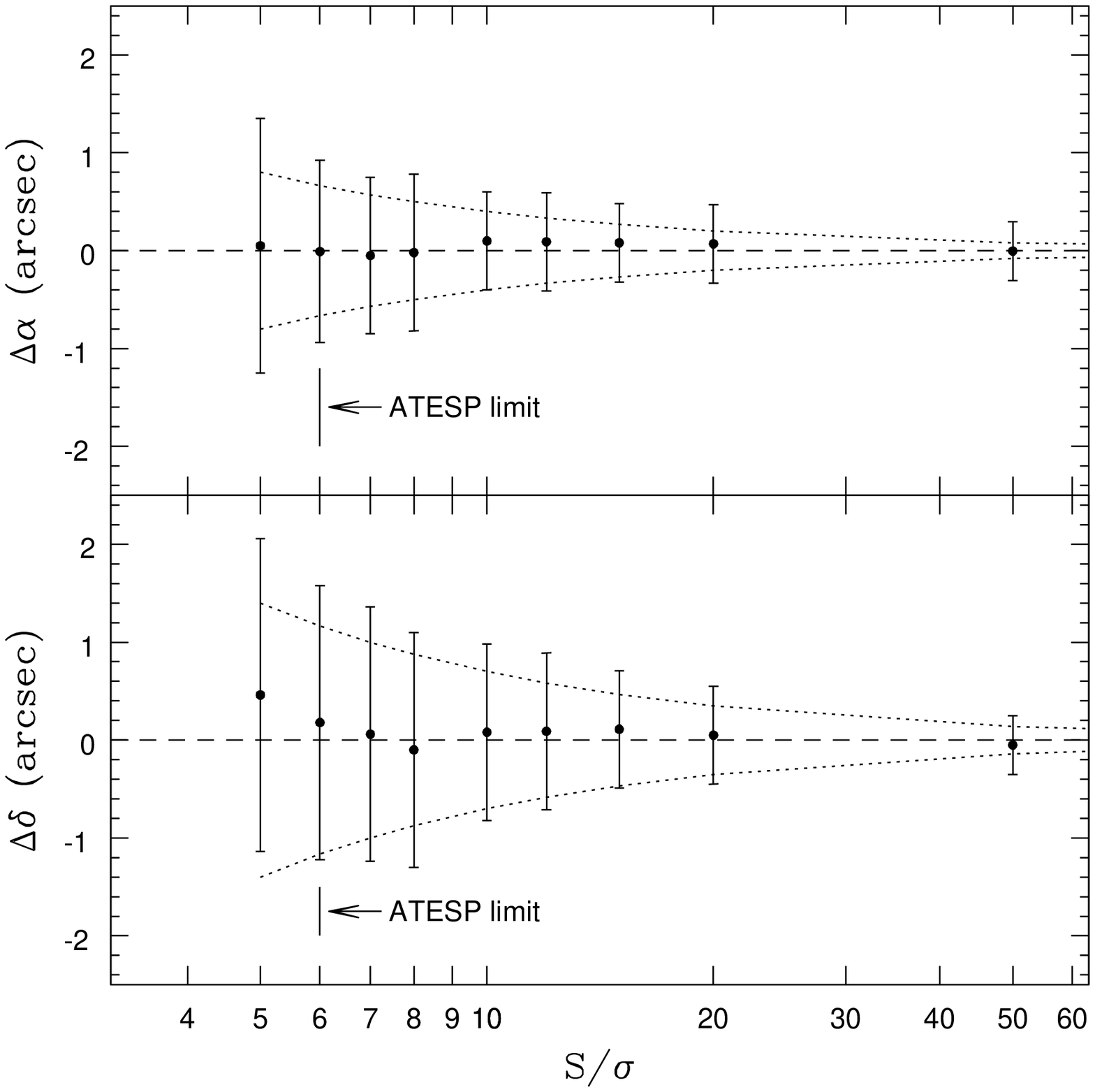}}
\hfill
\caption{Position internal accuracy for point sources. Top panel: 
right ascension. Bottom panel: declination.
Mean and standard deviation for the output-input (IMSAD-injected)
difference, as a function of flux. Expected 
values (see text) are also plotted for both the mean (dashed line) and the rms 
(dotted lines). The flux density cut-off chosen for the ATESP catalogue is
indicated by the vertical solid line.
\label{fig-possim}}
\end{figure}

The positional accuracy for point sources is shown in Fig.~\ref{fig-possim},
where we plot the difference ($\Delta \alpha$ and $\Delta \delta$) between 
the position found by 
IMSAD (output) and the injected one (input), as a function of flux. 
No systematic effects are present and the rms values are in agreement with 
the ones expected for point sources (Condon \cite{Condon98}, 
for details see Appendix~\ref{sec-appa}):
\begin{eqnarray} 
\sigma_{\alpha} & \simeq & \frac{b_{\rm min}}{2}\;\left(\frac{S_{\rm peak}}{\sigma}\right)^{-1} \label{eq-raerr}\\
\sigma_{\delta} & \simeq & \frac{b_{\rm maj}}{2}\;\left(\frac{S_{\rm peak}}{\sigma}\right)^{-1} \label{eq-decerr}
\end{eqnarray}
where we have assumed $b_{\rm min}=8\arcsec$ and $b_{\rm maj}=14\arcsec$, 
the average synthesized beam values of the ATESP survey (see dotted lines in 
Fig.~\ref{fig-possim}). The positional accuracy of ATESP sources is therefore
$\sim 1\arcsec$ at the limit of the survey ($6\sigma$), decreasing to $\sim
0.5\arcsec$ at $12\sigma$ ($\sim 1$ mJy) and to $\sim 0.1\arcsec$ at 
50$\sigma$.  

\subsection{Systematic Effects}\label{sec-corr}

\begin{figure}[t]
\resizebox{\hsize}{!}{\includegraphics{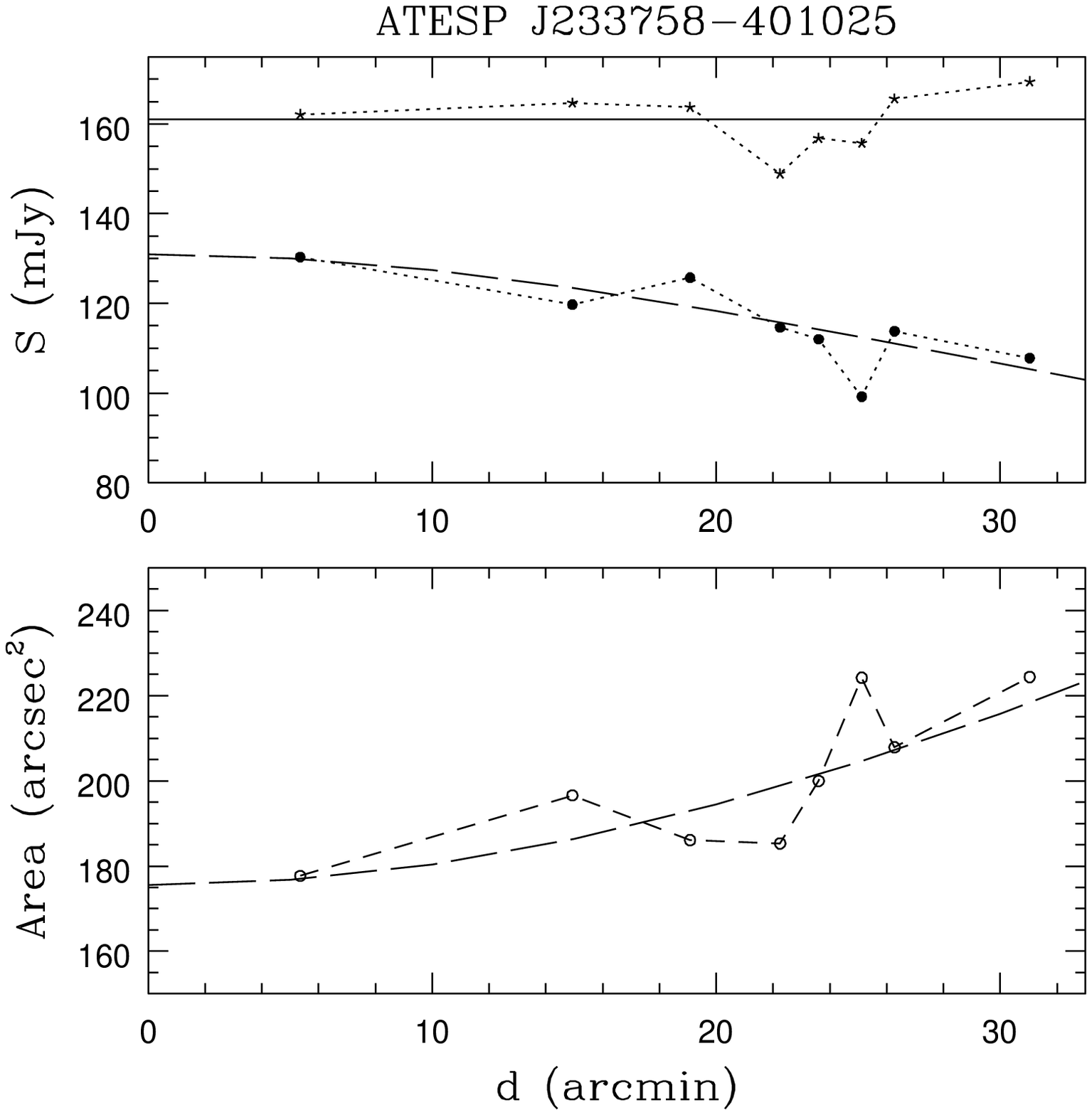}}
\hfill
\caption{Bandwidth smearing in ATESP single fields calibrated using 
source ATESP J233758-401025. The measured peak flux density
decreases going to larger distances (Top, full circles) while the measured
source area increases correspondingly (Bottom, empty circles). 
As a consequence the integrated flux density remains constant (Top, stars).
The curve describing the peak flux density -- distance can be used to 
describe the area -- distance relation as well. 
\label{fig-smear1}}
\end{figure}

\begin{figure}[t]
\resizebox{\hsize}{!}{\includegraphics{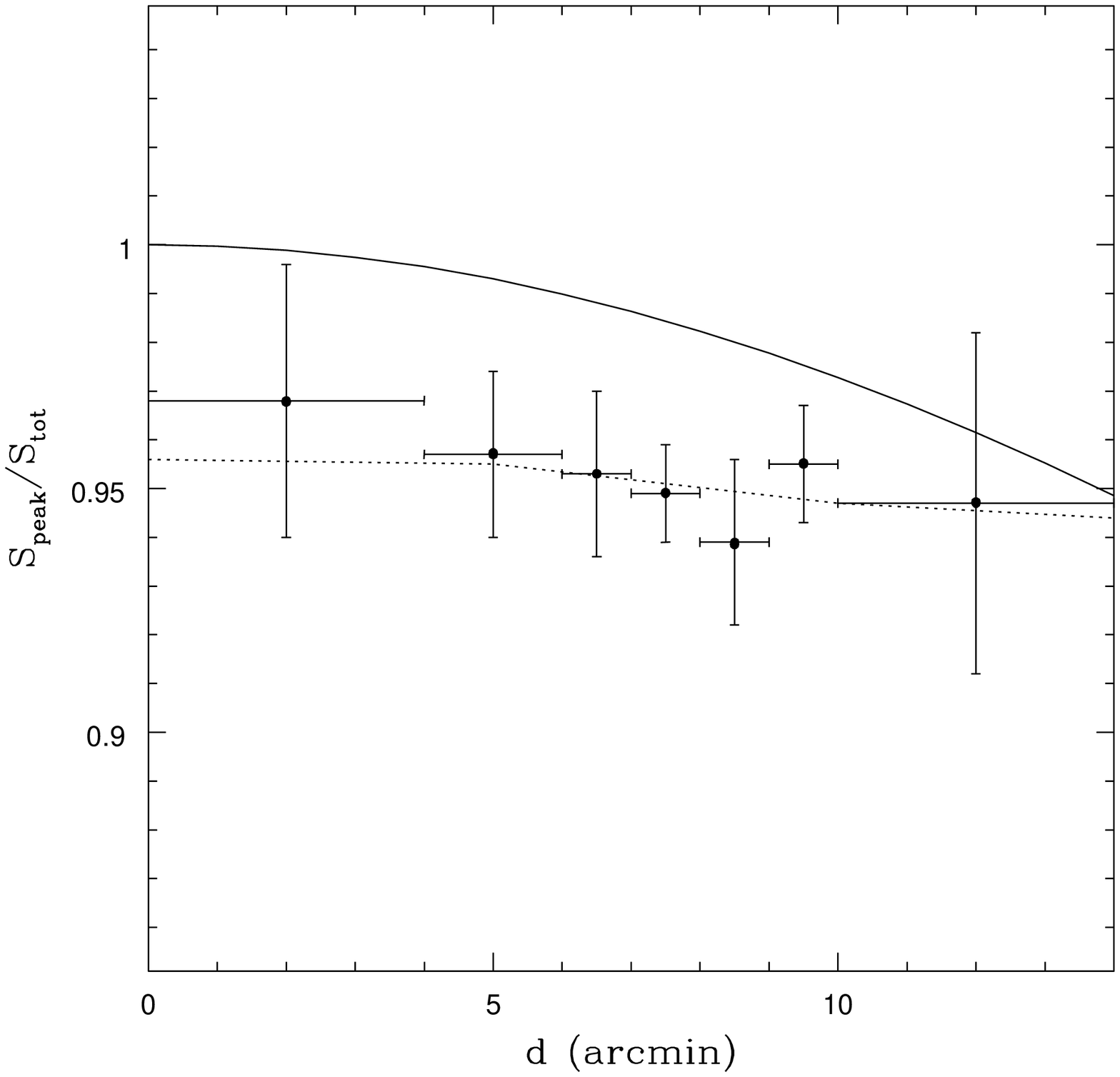}}
\hfill
\caption{Bandwidth smearing in ATESP mosaics.
Average $S_{\rm peak}/S_{\rm total}$ ratios obtained by 
summing all the unresolved ATESP sources brighter 
than $2$ mJy in different $d_{\rm min}$ intervals (full dots). 
$d_{\rm min}$ is defined as the distance to the closest field center. 
Also displayed are radial smearing expected in ATESP single fields 
(solid line) and in ATESP mosaics (dotted line). 
\label{fig-smear2}}
\end{figure}

Two systematic effects are to be taken into account when dealing with ATESP
flux densities, the clean bias and the bandwidth smearing effect. 
Clean bias has been extensively discussed in paper I of this series  
(see also Appendix~\ref{sec-appb} at the end of this paper). It is
responsible for flux density under-estimations of the order of $10-20\%$
at the lowest flux levels ($S<10\sigma$) and gradually disappears going to 
higher fluxes (no effect for $S\geq 50-100\sigma$). \\
The effect of bandwidth smearing is well-known. It 
reduces the peak flux density of a source, correspondingly increasing the 
source size in radial direction.
Integrated flux densities are therefore not affected. \\
The bandwidth smearing effect increases with the angular distance ($d$) 
from the the pointing center of phase and depends on the passband width
($\Delta\nu$), the observing frequency ($\nu$) and the synthesized 
beam FWHM width ($\theta_b$). The particular functional form that describes 
the bandwidth smearing is determined by the beam and the passband shapes.
It can be demonstrated, though, that the results obtained are not critically 
dependent on the particular functional form adopted (e.g. 
Bridle \& Schwab~\cite{Bridle89}). \\
In the simplest case of Gaussian beam and passband shapes, the 
bandwidth smearing effect can be described by the equation 
(see Eq.~12 in Condon et al.~\cite{Condonetal98}):
\begin{equation}\label{eq-sm}
\frac{S_{\rm peak}}{S^0_{\rm peak}} = \frac{1}{\sqrt{1 + \frac{2 \ln 2}{3}
\left( \frac{\Delta \nu }{\nu}\frac{d}{\theta_b} \right)^2 } }
\end{equation}
where the $\frac{S_{\rm peak}}{S^0_{\rm peak}}$ ratio represents the 
attenuation
on peak flux densities by respect to the unsmeared ($d=0$) source peak 
value. \\ 
We have estimated the actual smearing radial attenuation on
ATESP single fields, by measuring $S_{\rm peak}$ for a strong source 
(ATESP J233758-401025) detected in eight contiguous ATESP fields (corrected
for the primary beam attenuation) at 
increasing distance from the field center (full circles in 
Fig.~\ref{fig-smear1}, top panel). The data were then fitted using 
Eq.~\ref{eq-sm} by setting $\nu = 1.4$ GHz and 
$\theta_b \simeq 11\arcsec$ (from ($b_{\rm maj} + b_{\rm min})/2 = 
(14\arcsec +8\arcsec)/2$), as for the ATESP data. The best fit 
(dashed line in Fig.~\ref{fig-smear1}, top panel) gives 
$S^0_{\rm peak} = 131$ mJy and an effective 
bandwidth $\Delta \nu = 9$ MHz (in very good agreement with the nominal 
channel width $\Delta \nu = 8$ MHz (see Sect.~5.2 in paper I). 
As expected, the measured integrated flux density (stars in the same plot) 
remains constant. \\
The mosaicing technique consists in a weighted linear combination of all the 
single fields in a larger mosaiced image (see Eq.~1 in paper I). 
This means that, given single fields of size $1800\times 1800$ pixels,
source flux measures at distances as large as 
$\simeq 35\arcmin$ from field centers are still used to produce the final 
mosaic (even if with small weights). As a consequence, 
the radial dependence of bandwidth smearing tends to cancel out. \\
For instance, since ATESP pointings are organized in a $20\arcmin$ 
spacing rectangular grid, a source located at the center of phase of one 
field ($d=0$) is measured also   
at $d=20\arcmin$ in the 4 contiguous E, W, S and N fields 
and at  $d=20\cdot \sqrt{2} \simeq 28\arcmin$ in the other 4 \emph{diagonally} 
contiguous fields. Using Eq.~(1) of paper I, we can estimate a 4\%  
smearing attenuation for the mosaic peak flux of that source. In the same way
we can estimate indicative values for mosaic smearing attenuations as a 
function of $d_{\rm min}$, defined as the distance to the closest field center
(see dotted line in Fig.~\ref{fig-smear2}). 
We notice that actual attenuations vary from source to source depending on 
the actual position of the source in the mosaic. \\
From Fig.~\ref{fig-smear2} we can see that at small $d_{\rm min}$ mosaic 
smearing is much worse than single field's one (indicated by the solid line).
The discrepancy becomes smaller going to larger distances and disappears 
at $d_{\rm min}\simeq 14\arcmin$, which represents 
the maximum distance to the closest field center for ATESP sources.
This maximum $d_{\rm min}$ value gives an upper limit of $\sim 6\%$ to mosaic
smearing attenuations. \\
The expected mosaic attenuations have been compared to the ones obtained
directly estimating the smearing from the source catalogue. 
As already noticed (Sect.~\ref{sec-deconv}), a ratio
$S_{\rm peak}/S_{\rm total}< 1$, is purely determined, in case of point 
sources and in absence of flux measurement errors,  
by the bandwidth smearing effect, which systematically attenuates the source 
peak flux, leaving the integrated flux unchanged. \\ 
We have then considered all the unresolved ($\Theta_{\rm maj}=0$) ATESP
sources with $S_{\rm peak}>2$ mJy and we have plotted the average values of 
the $S_{\rm peak}/S_{\rm total}$ ratio in different distance intervals
(full dots in Fig.~\ref{fig-smear2}). The 2 mJy threshold ($\sim 25\sigma$)
was chosen in order to find a compromise between statistics and flux 
measure accuracy. 
The average flux ratios plotted are in very good agreement with the expected 
ones, especially when considering that the most reliable measures are the 
intermediate distance ones, where a larger number of sources can be summed.
In general we can conclude that on average smearing attenuations are 
$\sim 5\%$ and do not depend on the actual position of the source in the 
mosaics. This result also confirms the $5\%$ estimate drawn from 
Fig.~\ref{fig-stspratio}. 
We finally point out that smearing will affect to some extent also source
sizes and source coordinates.  

\subsection{Comparison with External Data}\label{sec-errcal}

To determine the quality of ATESP source parameters we have made comparisons
with external data.
Unfortunately in the region covered by the ATESP, there are not many 1.4 GHz
data available. The only existing 1.4 GHz radio survey is the NVSS 
(Condon et al. \cite{Condonetal98}), which covers about half of the region 
covered by the 
ATESP survey ($\delta > -40\degr$) with a point source detection 
limit $S_{lim}\sim 2.5$ mJy. \\
The NVSS has a poor spatial resolution 
($45\arcsec$ FWHM beam width) compared to ATESP and this introduces large 
uncertainties in the comparison, especially for astrometry. 
To test the positional accuracy we have therefore used data at other 
frequencies as well. In particular we have used VLBI 
sources extracted from the list of the standard calibrators at the ATCA 
and the catalogue of PMN compact sources with measured ATCA positions 
(Wright et al. \cite{Wright97}). 

\begin{figure}[t]
\resizebox{\hsize}{!}{\includegraphics{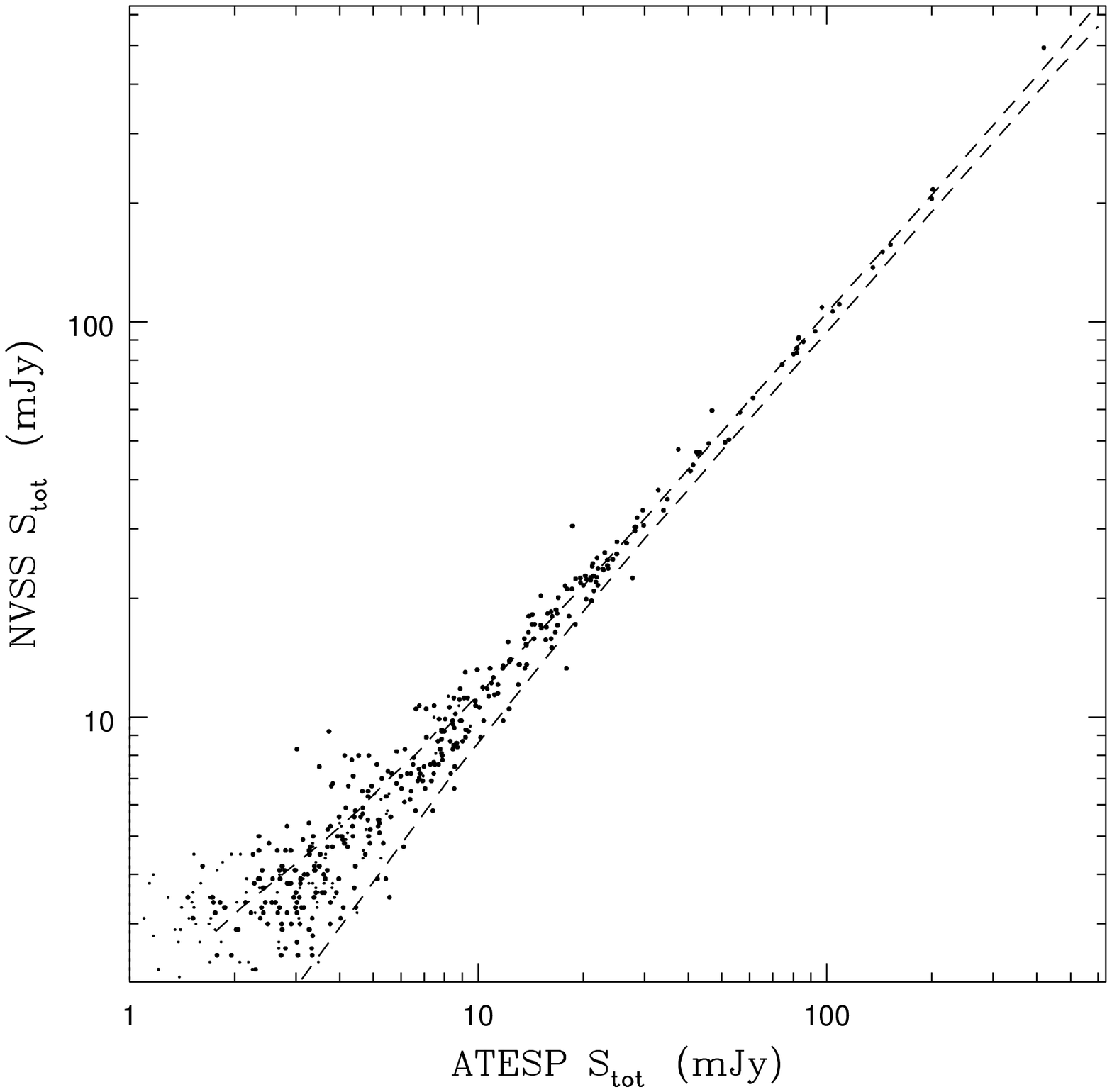}}
\hfill
\caption{Comparison of NVSS with ATESP flux densities. Dashed lines 
show the 90\% confidence limits in the flux measure. 
\label{fig-nvssatflux}}
\end{figure}
\begin{figure}[t]
\resizebox{\hsize}{!}{\includegraphics{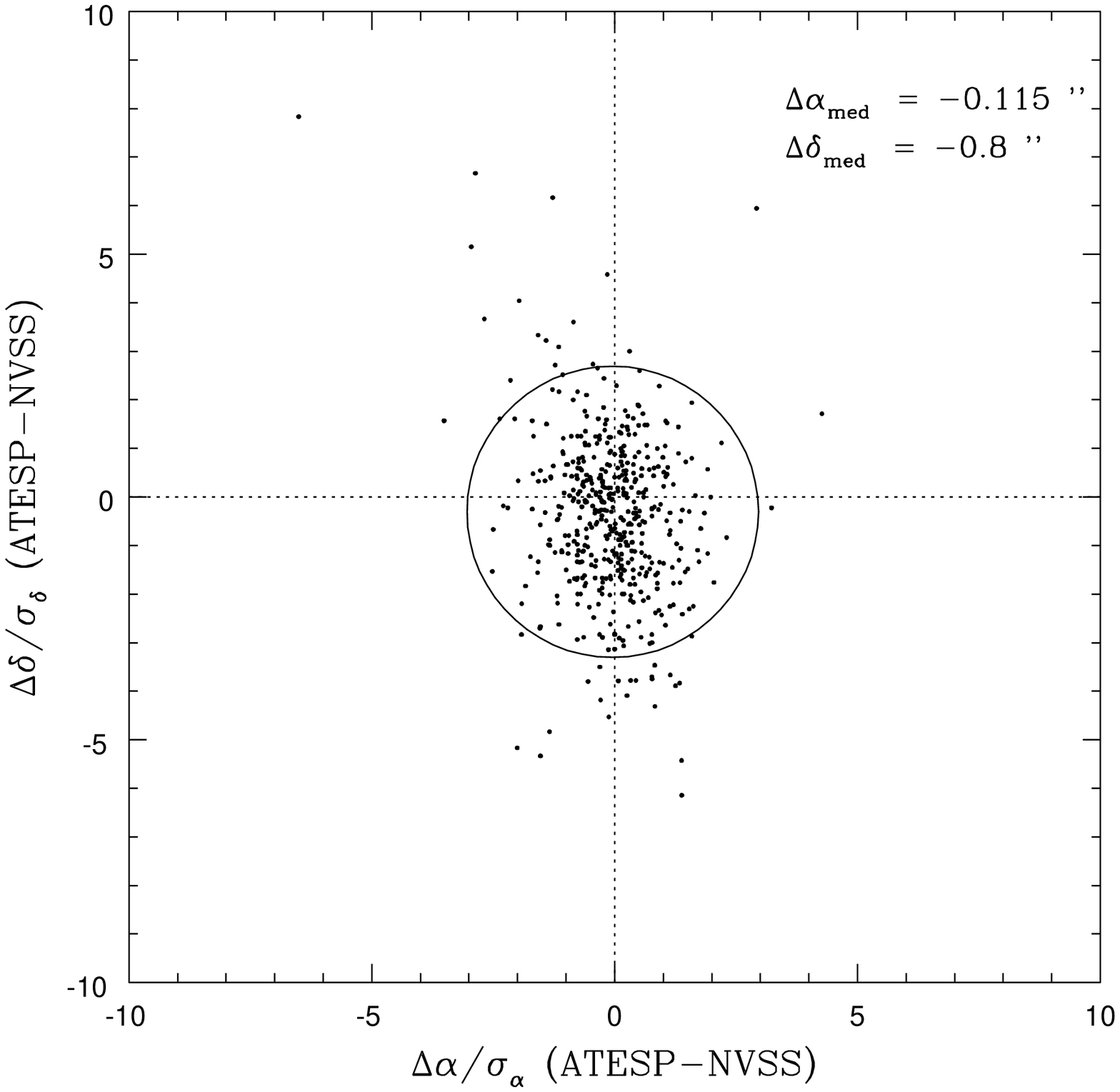}}
\hfill
\caption{Result of the cross-identification between ATESP and NVSS 
catalogues in the overlapping region. Only isolated ATESP sources have
been considered (see Sect.~\ref{sec-phot}). The position offsets ($\Delta\alpha$
and $\Delta\delta$ are expressed in terms of their rms uncertainties
($\sigma_{\alpha}$ and $\sigma_{\delta}$). The $3\sigma$ confidence 
error circle is also shown.
\label{fig-nvssatpos}}
\end{figure}

\subsubsection{Flux Densities}\label{sec-phot}

In order to estimate the quality of the ATESP flux densities we have compared 
ATESP with NVSS. To minimize the 
uncertainties due to the much poorer NVSS resolution we should in principle 
consider only point-like ATESP sources. Nevertheless,
in order to increase the statistics at high fluxes ($S>10$ mJy), 
we decided to include extended ATESP sources as well. Another source of 
uncertainty in the comparison is due to the fact that in the NVSS 
distinct sources closer together than $50\arcsec$ will be only marginally 
separated.
To avoid this problem we have  restricted the comparison to bright 
($S_{\rm peak}>1$ mJy)
\emph{isolated} ATESP sources: we have discarded all 
multi-component sources (as defined in Sect.~\ref{sec-mult}) and all 
single-component sources whose nearest neighbor is at a distance 
$\leq 100\arcsec$ (as in the comparison between the FIRST and the NVSS
by White et al. \cite{White97}). 
We have not used isolated ATESP sources fainter 
than 1 mJy because we have noticed that there are several cases where  
NVSS point sources are resolved in two distinct objects in the ATESP 
images, only one being listed in the ATESP catalogue 
(i.e. the other one has $S_{\rm peak} <6\sigma$).\\
In Fig.~\ref{fig-nvssatflux} we have plotted the NVSS against
the ATESP flux density for the 443 $S_{\rm peak} > 1$ mJy ATESP sources identified
(sources within the $3\sigma$ confidence circle in Fig.~\ref{fig-nvssatpos}). 
We have used integrated fluxes
for the sources which appear extended at the ATESP resolution (full circles) 
and peak fluxes for the unresolved ones (dots). Also indicated are the 
$90\%$ confidence limits (dashed curves), drawn taking into account 
both the NVSS and the ATESP errors in the flux measure. In drawing the upper
line we have also taken into account an average correction for the 
systematic under-estimation of ATESP fluxes due to clean bias and bandwidth 
smearing (see Sect.~\ref{sec-corr}). 
The provided NVSS fluxes are already corrected for any systematic effect. \\
The scatter plot shows that at high fluxes the ATESP flux scale 
agrees with the NVSS one within a few percents ($\leq 3\%$). This gives an
upper limit to calibration errors and/or resolution effects at high fluxes.
Going to fainter fluxes the discrepancies between the ATESP and the 
NVSS fluxes become larger, reaching deviations as high as a factor 
of 2 at the faintest levels. ATESP fluxes tend to be lower than NVSS fluxes. 
This could be, at least partly, due to resolution effects. 
Such effects can be estimated 
from the comparison between NVSS and ATESP sources itself
and from theoretical considerations.
Assuming the source integral angular size distribution provided by Windhorst 
et al. (\cite{Windhorst90}) we have that at the NVSS limit 
($S\simeq 2.5$ mJy) about 40\%
of the sources can be resolved by the ATESP 
synthesized beam (intrinsic angular sizes $\geq 4\arcsec$). This fraction 
goes up to $50\%$ at $S\simeq 10$ mJy.\\
On the other hand,
we point out that close to the NVSS catalogue cut-off, incompleteness could 
affect NVSS fluxes. 
For instance we have noticed that below 5 mJy, there are several cases where
the flux given in the NVSS catalogue is overestimated with respect to the one 
measured in the NVSS image (even taking into account the applied corrections).

\subsubsection{Astrometry}\label{sec-astr}

The region covered by the ATESP survey contains two VLBI sources
from the ATCA calibrator catalogue: 2227-399 and 0104-408.
These sources were not used to calibrate our data and therefore provide
an independent check of our ATCA positions. The offset between ATESP and VLBI 
positions (ATESP--VLBI) for the first and the second source respectively are:
$\Delta\alpha = -0.277\arcsec$ and $-0.023\arcsec$; $\Delta\delta = 
+0.239\arcsec$ and $-0.172\arcsec$. Such offsets indicate that the 
uncertainty in the astrometry should be within a fraction 
of arcsec. Obviously, we cannot exclude
the presence of systematic effects, but
an analysis of the ATESP--NVSS positional offsets gives
$\Delta\alpha_{\rm med}=-0.115\arcsec$ and $\Delta\delta_{\rm med}=-0.8\arcsec$
(see Fig.~\ref{fig-nvssatpos}). \\
A more precise comparison could be obtained by using the PMN sources 
with ATCA position measurements available. 
Unfortunately the number of such PMN sources in the region covered
by the ATESP survey is very small: we found only 12 identifications.
Using the 4.8 GHz positions for the PMN sources, we derived (ATESP--ATPMN) 
$\Delta\alpha_{\rm med}=-0.115\arcsec$ and $\Delta\delta_{\rm med}=-0.3\arcsec$. \\
We can conclude that all comparisons give consistent results and 
that the astrometry is accurate within a small fraction of an arcsec. Also 
systematic offsets, if present, should be very small. 

\begin{acknowledgements}
We acknowledge R. Fanti for reading and commenting on an 
earlier version of this manuscript. \\
The authors aknowledge the Italian Ministry for University and Scientific
Research (MURST) for partial financial support (grant Cofin 98-02-32). 
This project was undertaken under the CSIRO/CNR Collaboration programme.
The Australia Telescope is funded by the Commonwealth of Australia for 
operation as a National Facility managed by CSIRO.

\end{acknowledgements}

\appendix

\section{Master Equations for Error Derivation}\label{sec-appa}

As discussed in Sect.~\ref{sec-errors}, internal errors for the ATESP 
source parameters are well described by  Condon (\cite{Condon97}) 
equations of error 
propagation derived for two-dimensional elliptical Gaussian fits in presence 
of Gaussian noise. In order to get the total rms error on each parameter, 
a calibration term should be quadratically added.
Using Eqs.~(21) and (41) in Condon (\cite{Condon97}), total percentage 
errors for flux densities ($\frac{\sigma(S_{\rm peak})}{S_{\rm peak}}$) 
and fitted axes ($\frac{\sigma(\theta_{\rm maj})}{\theta_{\rm maj}}$, 
$\frac{\sigma(\theta_{\rm min})}{\theta_{\rm min}}$) can be calculated
from:
\begin{equation}\label{eq-condon}
\sqrt{\frac{2}{\rho^2} + \epsilon^2}
\end{equation}
where $\epsilon$ is the calibration term  and the effective signal-to-noise 
ratio, $\rho$, is given by:
\begin{equation}\label{eq-rho}
\rho^2 = \frac{\theta_{\rm maj} \theta_{\rm min}}{4 \theta_{\rm N}^2} 
\left[ 1 + \left(\frac{ \theta_{\rm N}}{\theta_{\rm maj}}\right)^2
\right]^{\alpha_{\rm M}} 
\left[ 1 + \left(\frac{ \theta_{\rm N}}{\theta_{\rm min}}\right)^2 
\right]^{\alpha_{\rm m}} \frac{S_{\rm peak}^2}{\sigma^2} 
\end{equation}  
where $\sigma$ is the image noise ($\sim 79$ $\mu$Jy on average for ATESP 
images), 
$\theta_{\rm N}$ is the FWHM of the Gaussian correlation length of the 
image noise (assumed $\simeq$ FWHM of the synthesized  beam) and the 
exponents are:
\begin{eqnarray} 
\alpha_{\rm M} = 3/2 \; \; {\rm and} &  \alpha_{\rm m} = 3/2 & {\rm for} \;
\; \sigma(S_{\rm peak}) \\
\alpha_{\rm M} = 5/2 \; \; {\rm and} &  \alpha_{\rm m} = 1/2 & {\rm for} \;
\; \sigma(\theta_{\rm maj}) \\ 
\alpha_{\rm M} = 1/2 \; \; {\rm and} &  \alpha_{\rm m} = 5/2 & {\rm for} \;
\; \sigma(\theta_{\rm min}) \;  .
\end{eqnarray}
Similar equations hold for position rms errors (Condon~\cite{Condon97}, 
Condon et al.~\cite{Condonetal98}):
\begin{eqnarray}
\sigma^2(\alpha) = & \epsilon^2_{\alpha} + \sigma^2(x_0) 
\sin^2({\rm P.A.}) + \sigma^2(y_0) \cos^2({\rm P.A.})  \label{eq-condonposa}\\
\sigma^2(\delta) = & \epsilon^2_{\delta} + \sigma^2(x_0) \cos^2({\rm P.A.}) 
+ \sigma^2(y_0) \sin^2({\rm P.A.}) \label{eq-condonposb}
\end{eqnarray}
where $\sigma^2(x_0) =\sigma^2(\theta_{\rm maj})/8\ln 2$ and 
$\sigma^2(y_0) =\sigma^2(\theta_{\rm min})/8\ln 2$ represent the rms lengths
of the major and minor axes respectively. \\
Calibration terms are in general estimated from comparison with 
consistent external data of better accuracy than the one tested. 
Unfortunately there are no such data available in the region of sky covered 
by the ATESP survey. Nevertheless, from our typical flux and phase 
calibration errors, we estimate calibration terms of about $5-10\%$ for both 
flux densities and source sizes. \\
As a caveat we remind (see discussion in Paper I) that the 500 m baseline 
cutoff applied to our data makes the ATESP survey progressively insensitive 
to sources larger than $30\arcsec$: assuming a Gaussian shape, only $50\%$
of the flux for a $30\arcsec$ large source would appear in the ATESP images.
It is important to have this in mind when dealing with flux densities and 
source sizes of the largest ATESP sources. \\
Right ascension and declination calibration terms have been estimated from 
the astrometry results reported in Sect.~\ref{sec-astr}. As already discussed,
the ATESP astrometry can be considered accurate within a small fraction of an
arcsec, even though the scarcity of (accurate) external data available in the 
ATESP region makes it difficult to quantify this statement. Nevertheless 
from the rms dispersion of the median offsets found between ATESP and the 
external comparison samples (see Sect.~\ref{sec-astr}) we can tentatively   
estimate $\epsilon_{\alpha}\simeq 0.1\arcsec$ and 
$\epsilon_{\delta}\simeq 0.4\arcsec$.\\
It can be easily demonstrated that the master equations~(\ref{eq-condon}), 
(\ref{eq-condonposa}) and (\ref{eq-condonposb}) reduce to 
Eqs.~(\ref{eq-sperr})~$-$~(\ref{eq-decerr}) in Sect.~\ref{sec-errors} 
(where the calibration term is neglected) in the case of ATESP point sources 
($\theta_{\rm maj}\times \theta_{\rm min} \simeq 14\arcsec \times 8\arcsec$, 
P.A.$\simeq +2\degr$), assuming 
$\theta_{\rm N}\simeq 11\arcsec$ (or $\theta_{\rm N}^2 \sim \theta_{\rm maj}
\theta_{\rm min}$). \\
For a complete and detailed discussion of the error master equations 
of source parameters obtained through elliptical Gaussian fits we refer to 
Condon (\cite{Condon97}) and Condon et al. (\cite{Condonetal98}). 

\section{Flux Density Corrections for Systematic Effects}\label{sec-appb}

As already discussed in Sect.~\ref{sec-corr}, two systematic effects are to 
be taken into account when dealing with ATESP flux densities,
the clean bias and the bandwidth smearing effect. \\
The flux densities reported in the ATESP source catalogue are not corrected 
for such systematic effects. The corrected flux densities ($S^{\rm corr}$) can be 
computed as follows:
\begin{equation}\label{eq-fluxcorr}
S^{\rm corr}=\frac{S^{\rm meas}}{k \cdot [a \log{(S^{\rm corr}/\sigma)} + b]} 
\end{equation} 
where $S^{\rm meas}$ is the flux actually measured in the ATESP images (the 
one reported in the source catalogue). The parameter $k$ 
represents the smearing correction. It is set equal to 1 (i.e. no correction) 
when the equation is applied to integrated flux densities and $< 1$ when 
dealing with peak flux densities. From the analysis reported in 
Sect.~\ref{sec-corr} we suggest to set $k=0.95$ ($5\%$ smearing effect). \\
The clean bias correction is taken into account by the term 
in the square brackets. As discussed in paper I, Sect.~5.3, the importance
of the clean bias effect varies from mosaic to mosaic depending on the 
average number of clean components (cc's). In particular we derived the values 
for the parameters $a$ and $b$ in three different mosaics representing
the case of low (fld34to40, 1616 cc's), intermediate (fld44to50, 2377 cc's) 
and high (fld69to75, 3119 cc's) average number of cc's 
(see Table~4 of paper I). \\
In correcting the source fluxes for the clean bias, 
we suggest to set $(a,b)=(0.09,0.85)$ whenever
the mosaic average number of cc's is $< 2000$ (low--cc's case); 
$(a,b)=(0.13,0.75)$ whenever the mosaic cc's average number is between 2000 and
3000 (intermediate--cc's case); $(a,b)=(0.16,0.67)$ whenever the mosaic cc's 
average number exceeds 3000 (high--cc's case). The average number of clean 
components for each mosaic is reported in Table~\ref{tab-mospar}. 
In order to trace back the sources to the original mosaics, 
Table~\ref{tab-mospar} lists also the right ascension range covered by each 
mosaic (indicated by the r.a. of the first and the last source in that 
mosaic). \\
The clean bias is a function of the source signal-to-noise ratio
$S^{\rm corr}/\sigma$. 
Since the noise level is fairly uniform within each mosaic, it is possible 
to assume $\sigma$ equal to the mosaic average noise value ($\sigma_{fit}$ in 
Table~\ref{tab-mospar}, we refer to paper I for details on mosaic noise 
analysis and average noise value definition). 
\begin{table}[t]
\caption[]{Mosaic parameters. \label{tab-mospar}}
\begin{flushleft}
\begin{tabular}{lrcc}
\hline\hline\noalign{\smallskip}
\multicolumn{1}{c}{Mosaic}   & \multicolumn{1}{c}{cc's}   
& \multicolumn{1}{c}{$\sigma_{fit}$ ($\mu$Jy)}
 & \multicolumn{1}{c}{R.A. range} \\
\noalign{\smallskip}
\hline\noalign{\smallskip} 
 & & & \\
fld01to06 & 2033 &  78.7 & $22:32:35$ - $22:39:24$ \\
fld05to11 & 1796 &  77.8 & $22:39:28$ - $22:49:51$ \\
fld10to15 & 3104 &  88.1 & $22:49:54$ - $22:56:46$ \\
fld20to25 & 2823 &  83.0 & $23:31:30$ - $23:38:20$\\
fld24to30 & 2716 &  82.8 & $23:38:27$ - $23:48:51$ \\
fld29to35 & 2044 &  79.2 & $23:48:54$ - $23:55:43$ \\
fld34to40 & 1616 &  76.3 & $23:55:51$ - $00:06:15$ \\
fld39to45 & 2535 &  81.2 & $00:06:21$ - $00:13:10$ \\
fld44to50 & 2377 &  78.0 & $00:13:20$ - $00:23:35$ \\
fld49to55 & 2168 &  78.6 & $00:23:45$ - $00:30:35$ \\
fld54to60 & 2504 &  77.3 & $00:30:40$ - $00:41:01$ \\
fld59to65 & 2447 &  79.4 & $00:41:07$ - $00:47:57$ \\
fld64to70 & 1899 &  75.1 & $00:48:03$ - $00:58:29$ \\
fld69to75 & 3119 &  81.9 & $00:58:30$ - $01:05:22$ \\
fld74to80 & 2558 &  77.1 & $01:05:26$ - $01:15:50$ \\
fld79to84 & 1522 &  68.9 & $01:15:53$ - $01:22:50$ \\
 & & & \\
\noalign{\smallskip}
\hline\hline
\end{tabular}
\end{flushleft}
\end{table}


\begin{thebibliography}{}

\bibitem[1977]{Baarsetal77}
Baars J.W.M., Genzel R., Pauliny-Toth I.I.K., Witzel A., 1977, A\&AS 61, 99
\bibitem[1989]{Bridle89}
Bridle A.H., Schwab F.R., 1989. In: Perley et al. (eds.) Synthesis Imaging 
in Radio Astronomy, ASP Conf. Ser. 6, 247
\bibitem[1997]{Condon97}
Condon J.J., 1997, PASP 109, 166
\bibitem[1998]{Condon98}
Condon J.J., 1998. In: B.J. McLean et al. (eds.) Proc. IAU 
Symp. 179, New Horizons from Multi-Wavelength Sky Surveys, p.~19
\bibitem[1998]{Condonetal98}
Condon J.J., Cotton W.D., Greiser E.W., et al., 1998, AJ 115, 1693
\bibitem[2000]{Prandoni00}
Prandoni I., Gregorini L., Parma P., et al., 2000, A\&AS 146, 31 (Paper I)
\bibitem[1995]{Sault95}
Sault R.J., Killeen N., 1995, Miriad Users Guide
\bibitem[1997]{Vettolani97}
Vettolani G., Zucca E., Zamorani G., et al., 1997, A\&A 325, 954
\bibitem[1997]{White97}
White R.L., Becker R.H., Helfand D.J., Gregg M.D., 1997, ApJ 475, 479
\bibitem[1990]{Windhorst90}
Windhorst R.A., Mathis D., Neuschaefer L., 1990. In: Kron R.G. (ed.) 
Evolution of the Universe of Galaxies, ASP Conf. Ser. 10, 389
\bibitem[1997]{Wright97}
Wright A.E., Tasker N., McConnell D., et al., 1997, 
 http://www.parkes.atnf.csiro.au/databases/surveys/
\end{thebibliography}
\end{document}